\documentclass{nat} 

\usepackage{xcolor}
\usepackage{graphicx}
\usepackage{dcolumn}
\usepackage{bm}
\usepackage{color}
\usepackage[colorlinks=true,linkcolor=blue,urlcolor=blue,citecolor=blue,anchorcolor=blue]{hyperref}\usepackage{units}
\usepackage{hhline}
\usepackage{amsmath,amssymb}
\usepackage{float}
\usepackage{caption}
\usepackage{subcaption}

\newcommand{\Like}{\mathcal{L}}

\newcommand{\CSIRO}{Australia Telescope National Facility, CSIRO, Space and Astronomy, PO Box 76, Epping, NSW 1710, Australia.}
\newcommand{\Swin}{Centre for Astrophysics and Supercomputing, Swinburne University of Technology, PO Box 218, Hawthorn, VIC 3122, Australia.}
\newcommand{\OzGrav}{The ARC Centre of Excellence for Gravitational-wave Discovery (OzGrav).}
\newcommand{\Perdue}{Department of Physics and Astronomy, Purdue University, 525 Northwestern Avenue, West Lafayette, IN 7907-2036, USA.}
\newcommand{\USyd}{Sydney Institute for Astronomy (SIfA), School of Physics, The University of Sydney, Camperdown NSW 2006, Australia.}
\newcommand{\SARAO}{South African Radio Astronomy Observatory, 2 Fir Street, Black River Park, Observatory 7925, South Africa.}
\newcommand{\AstroThreeD}{The ARC Centre of Excellence for All Sky Astrophysics in 3 Dimensions (ASTRO3D).}
\newcommand{\ManUni}{Jodrell Bank Centre for Astrophysics, Department of Physics and Astronomy, The University of Manchester, Manchester M13 9PL, UK.}
\newcommand{\WSU}{School of Science, Western Sydney University, Locked Bag 1797, Penrith, NSW 2751, Australia.}
\newcommand{\ASTRON}{ASTRON, the Netherlands Institute for Radio Astronomy, Oude Hoogeveensedijk 4, NL-7991 PD Dwingeloo, The Netherlands.}
\newcommand{\ParkesObs}{Australia Telescope National Facility, CSIRO, Space and Astronomy, P.O. Box 276, Parkes, NSW 2870 Australia.}

\newcommand{\FAST}{CAS Key Laboratory of FAST, NAOC, Chinese Academy of Sciences, Beijing, China.}
\newcommand{\UChineseAcadamy}{University of Chinese Academy of Sciences, Beijing, China.}
\newcommand{\CAG}{Computational Astronomy Group, Zhejiang Laboratory, Hangzhou 311100, China.}

\title{Linear to circular conversion in the polarized radio emission of a magnetar}

\author
{Marcus~E.~Lower,$^{1}$
Simon~Johnston,$^{1}$
Maxim~Lyutikov,$^{2}$
Donald~B.~Melrose,$^{3}$
Ryan~M.~Shannon,$^{4,5}$
Patrick~Weltevrede,$^{6}$\\
Manisha~Caleb,$^{3,7}$
Fernando~Camilo,$^{8}$
Andrew~D.~Cameron,$^{4,5}$
Shi~Dai,$^{9}$
George~Hobbs,$^{1}$
Di~Li,$^{10,11,12}$
Kaustubh~M.~Rajwade,$^{6,13}$
John~E.~Reynolds,$^{1}$
John~M.~Sarkissian$^{14}$
and
Benjamin~W.~Stappers$^{6}$
\\
\\
\normalsize{$^{1}$\CSIRO{}}\\
\normalsize{$^{2}$\Perdue{}}\\
\normalsize{$^{3}$\USyd{}}\\
\normalsize{$^{4}$\Swin{}}\\
\normalsize{$^{5}$\OzGrav{}} \\
\normalsize{$^{6}$\ManUni{}}\\
\normalsize{$^{7}$\AstroThreeD{}}\\
\normalsize{$^{8}$\SARAO{}}\\
\normalsize{$^{9}$\WSU{}}\\
\normalsize{$^{10}$\FAST{}}\\
\normalsize{$^{11}$\UChineseAcadamy{}}\\
\normalsize{$^{12}$\CAG{}}\\
\normalsize{$^{13}$\ASTRON{}}\\
\normalsize{$^{14}$\ParkesObs{}}\\
}

\begin{document}
\maketitle

\begin{abstract}
Radio emission from magnetars provides a unique probe of the relativistic, magnetized plasma within the near-field environment of these ultra-magnetic neutron stars. 
The transmitted waves can undergo birefringent and dispersive propagation effects that result in frequency-dependent conversions of linear to circularly polarized radiation and vice-versa, thus necessitating classification when relating the measured polarization to the intrinsic properties of neutron star and fast radio burst (FRB) emission sites. 
We report the detection of such behavior in 0.7--4\,GHz observations of the P = 5.54\,s radio magnetar XTE~J1810$-$197 following its 2018 outburst. 
The phenomenon is restricted to a narrow range of pulse phase centered around the magnetic meridian.
Its temporal evolution is closely coupled to large-scale variations in magnetic topology that originate from either plastic motion of an active region on the magnetar surface or free precession of the neutron star crust. 
Our model of the effect deviates from simple theoretical expectations for radio waves propagating through a magnetized plasma. 
Birefringent self-coupling between the transmitted wave modes, line-of-sight variations in the magnetic field direction and differences in particle charge or energy distributions above the magnetic pole are explored as possible explanations.
We discuss potential links between the immediate magneto-ionic environments of magnetars and those of FRB progenitors.
\end{abstract}


Magnetars are a class of slowly rotating neutron star with implied surface magnetic-field strengths exceeding $10^{14}$\,G. 
The majority of magnetars are exclusively detected at X-ray and gamma-ray wavelengths (see ref.\cite{Olausen2014}; \url{https://www.physics.mcgill.ca/~pulsar/magnetar/main.html}). 
To date, only six have been found to emit radio pulses.
The magnetar XTE~J1810$-$197 was initially discovered following a high-energy outburst in 2003\cite{Ibrahim2004} with a peculiar flat-spectrum radio counterpart\cite{Halpern2005}. 
Followup observations revealed the magnetar emitted bright, highly polarized radio pulses\cite{Camilo2006}.
The radio emission displayed dramatic changes in pulse profile shape and intensity over time\cite{Camilo2007b, Kramer2007}.
Polarimetry of individual single pulses revealed the presence of low level circular polarization with an apparent frequency-dependent change in handedness\cite{Kramer2007}. 
This behavior was believed to be the product of the electromagnetic radiation propagating through a birefringent medium such as the highly-magnetized, relativistic electron-positron pair plasma within neutron star magnetospheres\cite{Goldreich1969}.
Polarization-dependent dispersion within such a plasma can result in an efficient conversion between the linearly and circularly polarized components of the transmitted radiation, an effect known as Faraday conversion (sometimes referred to as generalized Faraday rotation or the Cotton-Mouton/Voigt effect)\cite{Sazonov1969, Kennett1998}. 
This propagation effect has been invoked as a means for generating rotation-phase dependent variations in the linear and circular polarization of some pulsars\cite{Edwards2004, Noutsos2009, Ilie2019, Sobey2021}, and was recently suggested as the mechanism responsible for generating circular polarization within the eclipses of the pulsar binary system PSR~J1748$-$2446A\cite{Li2023}.
A similar linear-to-circular conversion can take place due to either fully or partially coherent averaging of linearly polarized emission modes that have undergone birefringent delays within neutron star magnetospheres\cite{Melrose1977, Petrova2001, Dyks2019}. 
Partially coherent mode mixing has been recently investigated as a means for explaining non-orthogonal jumps in linear polarization position angles and frequency-dependent curricular polarization variations in young pulsars\cite{Oswald2023}.
Recent advances in wide-bandwidth radio telescope instrumentation and signal processing systems have since allowed us to identify and characterize these effects and use them to probe the immediate plasma environment in and around the magnetospheres of these extreme objects.

%
\section*{Observations}

After more than a decade of quiescence, radio pulses were again detected from XTE~J1810$-$197 on MJD 58460, following a high-energy outburst that began sometime between MJD 58442--58448\cite{Levin2019, Gotthelf2019}.
We began a monitoring campaign of the magnetar using the CSIRO Parkes 64-m radio telescope (also known as \textit{Murriyang}) and 76-m Lovell Telescope at the Jodrell Bank Observatory (JBO).
Here we focus on the observations taken between MJD 58460--58512.
A total of ten observations were performed during this time period using the Parkes Ultra-Wideband Low (Parkes-UWL) receiver system, covering a broad frequency range extending from 704--4,032\,MHz.
At JBO, a set of 33 observations that were taken using a comparatively narrowband (512\,MHz) receiver system centered at 1,532\,MHz.
All observations were corrected for frequency-dependent dispersion and Faraday rotation introduced by propagation of the radio pulses through the interstellar medium.
Details of the observations and data processing procedure can be found in the Methods and Extended Data Table \ref{tab:obs}.

\section*{Results and analysis}

\begin{figure}[t!]
    \centering
    \includegraphics[width=\linewidth]{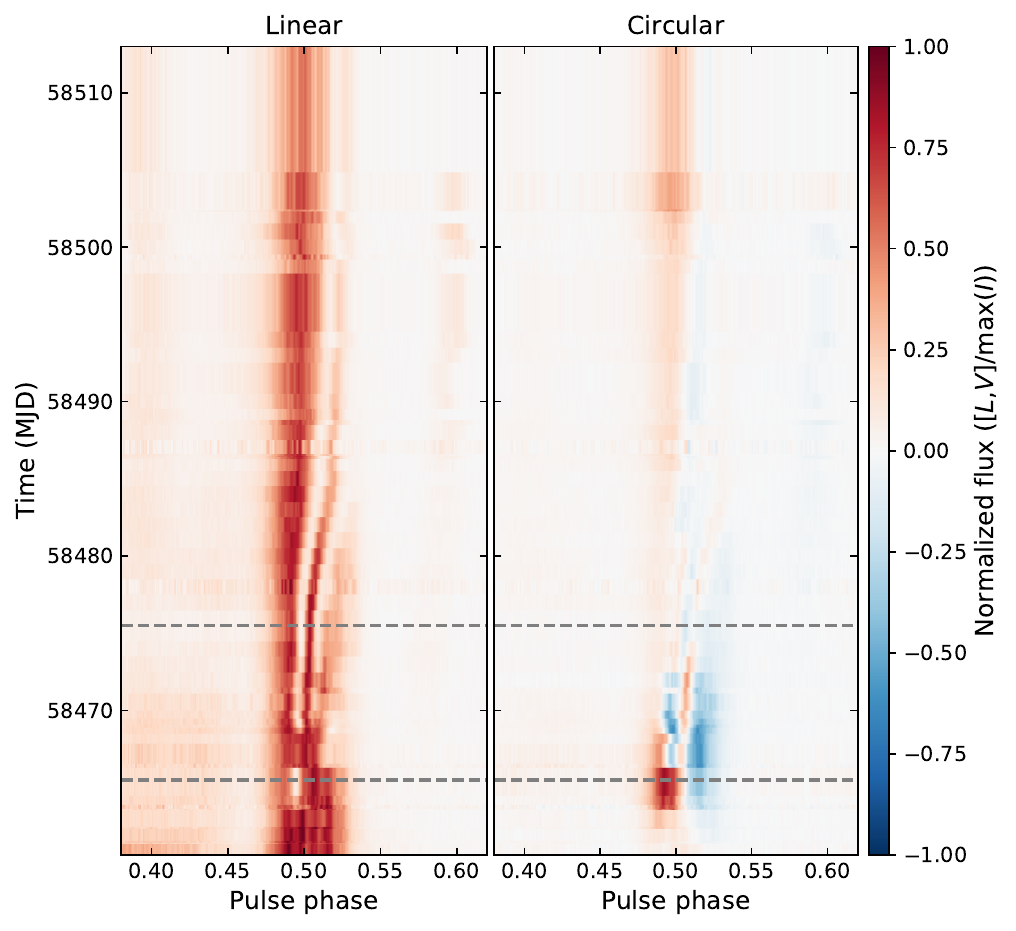}
    \caption{{\bf Polarization evolution of XTE~J1810$-$197 with time.} Linear and circular polarized emission detected by Lovell and Parkes observed at each epoch after averaging in frequency over 1,340--1,724\,MHz and 0.07--2.0\,hrs. These data are presented as functions of pulse phase and time. Red and blue colors correspond to positive and negative values respectively. The initial smooth linear variations across pulse phase and slowly increasing level of positive (right-handed) circular polarization undergo a dramatic change after 13 December 2018 (MJD 58465; lower horizontal line). Sudden reversals in circular handedness correspond to dips in linear intensity due to strong conversion effects. The circular intensity diminishes within $\sim 12$\,d of the 13 December 2018 event (MJD 58477; upper horizontal line). It was then followed by an apparent drift of the linear-to-circular conversion feature to increasingly later phases with time.}
    \label{fig:stokes_time}
\end{figure}

The radio emission from XTE~J1810$-$197 after its reactivation was comprised of narrow, millisecond-duration sub-pulses\cite{Dai2019, Maan2019}.
Averaging hundreds of rotations over time results in an average pulse profile comprised of three distinct, highly polarized components: a strong `central' component flanked by a wide leading component and a narrow trailing component (Extended Data Figs. \ref{fig:post_stamps} and \ref{fig:water_stamps}).
Simple power-law fits to the total intensity spectrum across the pulse profile revealed spectral indices ($\kappa$; see Methods for details), with predominantly negative values across the precursor and trailing components, whereas the central component displayed either a flat or inverted spectrum (Extended Data Fig.~\ref{fig:spec_index}).
The component-dependent spectral indices remained fairly stable over the first two months of the outburst, unlike much of the other radiative properties of the magnetar which displayed substantial temporal evolution.
In Fig.~\ref{fig:stokes_time} we highlight epoch-to-epoch changes in the frequency-averaged linear and circular polarization of XTE~J1810$-$197 normalized by the peak total intensity flux.
We observed remarkable variations in the linear and circular polarization of the central component as a function of pulse phase and time.
An initial increase in left- and right-hand circular polarization and simultaneous decrease in linear polarization in the leading half of the component were abruptly interrupted sometime between MJDs 58464--58466.
This coincided with the disappearance of an apparent quasi-periodic spacing between sub-structures in the central profile component\cite{Levin2019} and an inversion of the linear polarization position angle (PA) swing across the profile (Extended Data Fig.~\ref{fig:post_stamps}).
A second such PA swing inversion occurred between MJDs 58532--58343 but was not associated with polarization variations across the central component.

\begin{figure*}[t!]
    \centering
    \includegraphics[width=\linewidth]{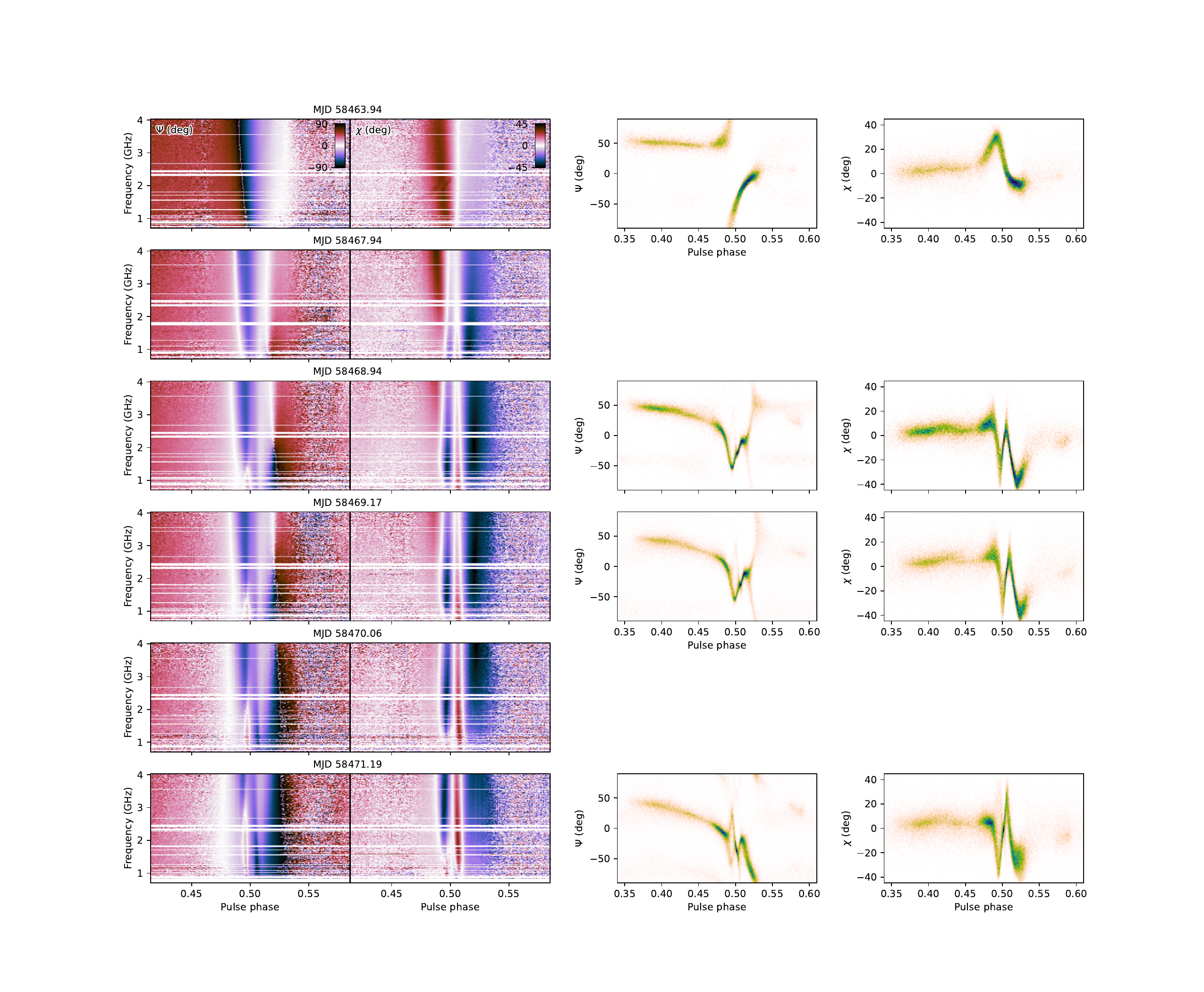}
    \caption{{\bf Evolution of the phase and frequency resolved polarization properties of XTE~J1810$-$197.} Left-hand panels show the linear polarization position angle ($\Psi$; left) and ellipticity angle ($\chi$; middle-left) for the Parkes-UWL observations as a function of pulse phase and observing frequency. Blank horizontal lines correspond to frequency channels that were excised due to contamination by radio-frequency interference. Right-hand panels display the frequency-averaged, single-pulse distributions of $\Psi$ (middle-right) and $\chi$ (right) for the four Parkes-UWL search-mode observations. Large frequency-dependent variations and branching paths in the single-pulse distributions caused by propagation effects in the magnetar magnetosphere are evident in observations from MJD 58467 onward.}
    \label{fig:pa_chi}
\end{figure*}

Following the initial PA-inversion the magnetar began to display clear frequency-dependent variations in its polarization properties.
This is highlighted in Fig.~\ref{fig:pa_chi} where we show the frequency and phase resolved position and ellipticity angles, $\Psi = 0.5\tan^{-1}(U/Q)$ and $\chi = 0.5\tan^{-1}(V/L)$, across the central profile component from MJD 58463--58471.
The right-hand panels of Fig.~\ref{fig:pa_chi} show the single pulse PA and ellipticity angle (EA) distributions measured on MJDs 58468, 58469 and 58471 gradually branch out along two separate paths at the same pulse phases where significant left-hand circular polarization was detected. 
Such rapidly diverging PA distributions are often attributed to radio emission from two orthogonally polarized modes (OPMs)\cite{Manchester1975}.
While there does appear to be a weak OPM in the precursor and trailing components, along with a clear OPM jump at pulse phases between 0.52--0.53 on MJDs 58468--58471, the branching PA distributions in question are inconsistent with switching between orthogonal emission states.
The separation and recombination of the PA swing distributions are substantially slower than the near-instantaneous transitions expected for OPMs (see ref.\cite{Karastergiou2005}).
Additionally, while the `orthogonal modes' are 90-deg apart from one another at pulse phases between 0.495 and 0.508, they are only offset by $\pm 45$\,deg from the nominal PA track.
This instead points to the polarization variations having arisen from some form of birefringent propagation effect either within the magnetosphere or near-field environment of XTE~J1810$-$197\cite{Petrova2001, Dyks2020}.

\begin{figure*}[t!]
    \centering
    \includegraphics[width=0.95\linewidth]{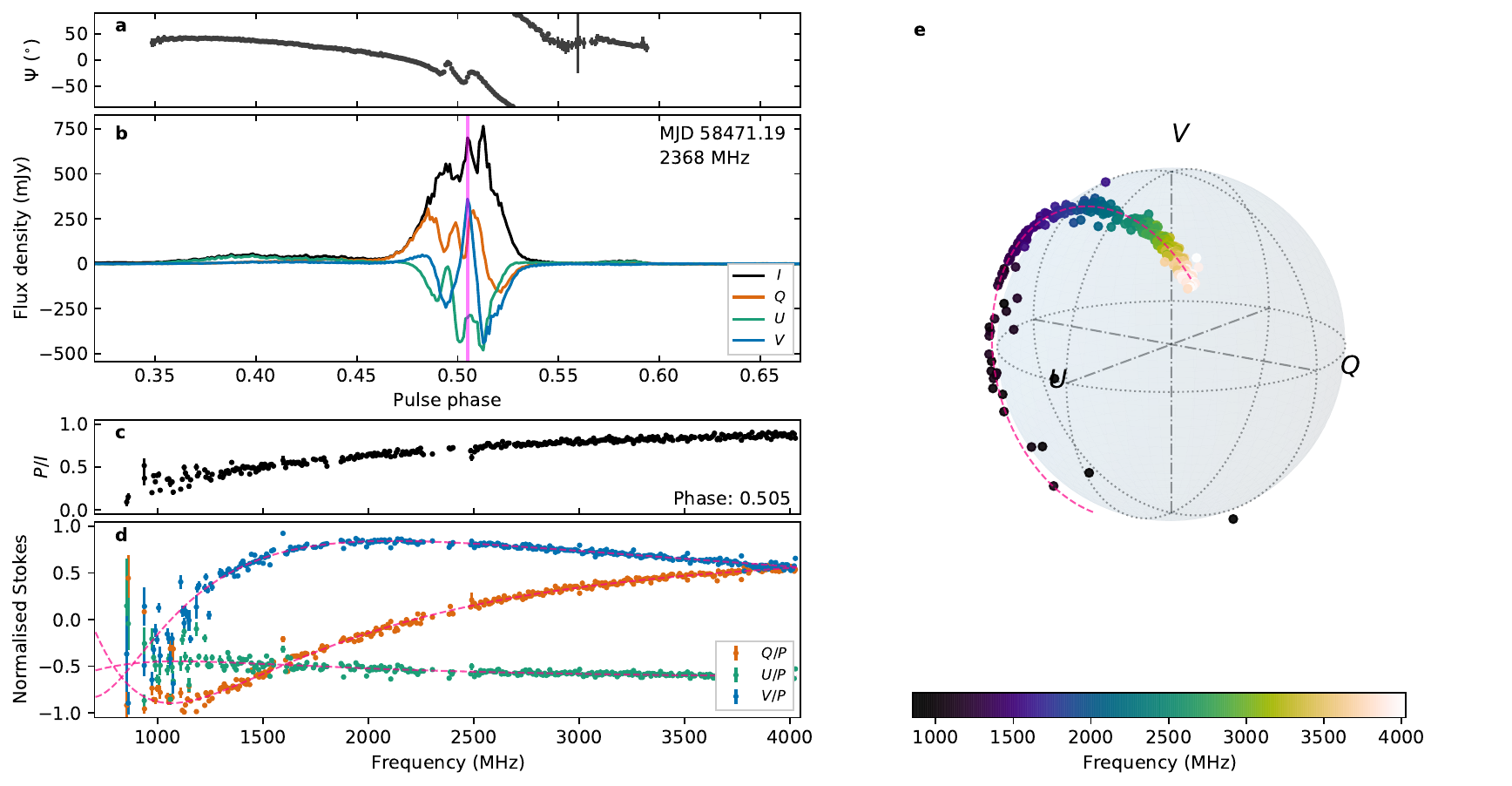}
    \caption{{\bf Example detection of linear-to-circular conversion in XTE~J1810$-$197.} Linear polarization position angle and 1-$\sigma$ uncertainties (a) along with the time- and frequency-averaged polarization profile (b) of XTE~J1810$-$197 detected on 19 December 2018, where the black, orange, green and blue lines respectively correspond to the total intensity ($I$), two linear polarizations ($Q$ and $U$) and the circular polarization ($V$). The vertical magenta line indicates the phase bin (0.505 in this plot) for which we plot the corresponding time-averaged total polarization fraction (c) and polarization spectra (d) where the error bars correspond to the off-pulse root-mean-square flux. In the latter, Stokes $Q$, $U$ and $V$ were normalized by total polarization ($P$), which traces out an clear frequency-dependent circle on the Poincar\'{e} sphere (e). The dashed magenta lines in panels c and d correspond to the median a-posteriori fit to the data. Animated versions of this figure that scan across pulse phase and include different epochs are depicted in Supplementary Videos 1-10.}
    \label{fig:181219_bin534}
\end{figure*}

\begin{figure*}[t!]
    \centering
    \includegraphics[width=\linewidth]{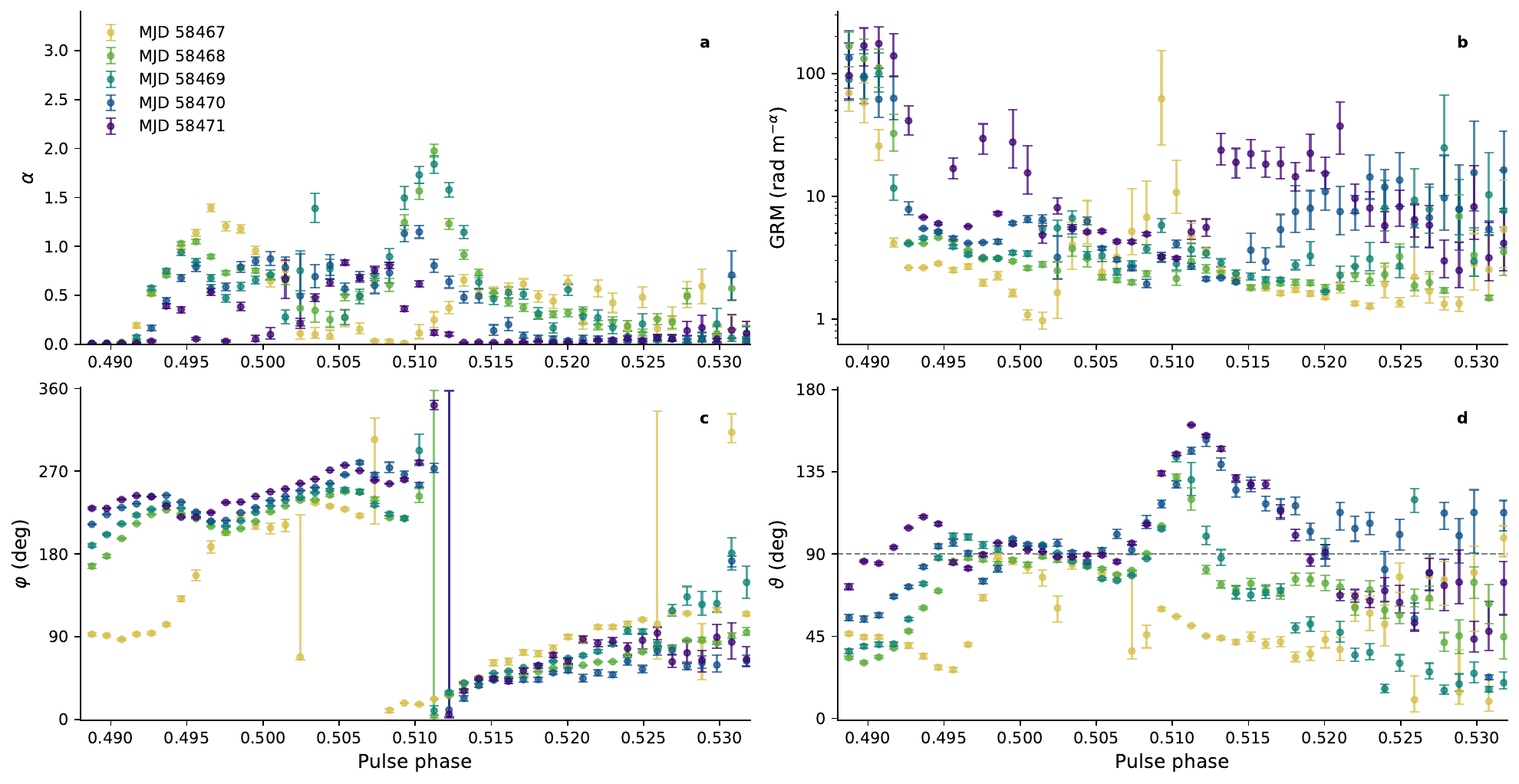}
    \caption{{\bf Recovered geometric parameters from the polarization conversion model for five Parkes-UWL epochs.} Median a-posteriori values (points) 68\% confidence intervals (error bars) for the spectral exponent (a), generalized rotation measure in units of rad\,m$^{-\alpha}$ (b), and the longitude (c) and latitude (d) of the conversion axis. Different colors represent points measured from different epochs between MJD 58467--58471, as indicated in the key located in the upper-left.}
    \label{fig:gfr_fits}
\end{figure*}

To understand what is driving the frequency-dependent polarization changes in XTE~J1810$-$197, we analyzed the Stokes spectra extracted from individual pulse phase bins within the central component.
In Fig.~\ref{fig:181219_bin534} we plot the polarization fraction and normalized Stokes $Q$, $U$ and $V$ parameters corresponding to the peak in right-hand circular polarization on MJD 58471.
All three Stokes parameters show strong variations that become more intense at lower observing frequencies, which coincides with a gradual decline in total polarization fraction.
We characterized these variations by fitting the polarization spectra in each phase bin with a phenomenological model\cite{Lower2021b} (see Methods for details).
The priors on the model parameters are listed in Table~\ref{tab:gfr_priors}.
Our median recovered values and 68\% confidence intervals are presented in Fig.~\ref{fig:gfr_fits} for the five observations where intense circular polarization was detected.
These include the wavelength dependence ($\alpha$), generalized rotation measure (GRM), rotation about the Stokes $V$ axis ($\varphi$) and tilt angle between the polarization vector rotation axis away from the right-hand Stokes $V$ pole ($\vartheta$).
Extended Data Fig.~\ref{fig:chi_sq} depicts the reduced chi-square statistic for 1,000 random draws per model spectrum.
All four model parameters vary substantially with pulse phase and observing epoch, indicating a strong line-of-sight dependence.
A significant degree of clustering around $\vartheta \approx 90$\,deg occurred at phases between 0.495--0.507 in most epochs.
Outside this region of pulse phase, the value of $\vartheta$ varies between $0$--$90$\,deg and $90$--$180$\,deg.
The wavelength dependence and GRM also vary with pulse phase and observing epoch, with $\alpha$ generally taking on values that are smaller than theoretical predictions for Faraday conversion\cite{Kennett1998, Gruzinov2019, Lyutikov2022}.
Some of this behavior can be attributed to only weak frequency-dependent variations occurring at phases earlier than 0.492 and later than 0.53.
However, this does not adequately explain the consistent measurements of $\alpha < 1$ or the near-zero wavelength dependency recovered around pulse phases of 0.5 or after 0.51 on MJDs 58470 and 58471, where the model fails to reconstruct the data below 1,300\,MHz (see Extended Data Fig.~\ref{fig:181219_bin504} and the animated versions of Fig.~\ref{fig:181219_bin534} in the Supplementary Materials).

We investigated whether rapid fluctuations in the propagating medium could explain the deviations away from a simple $\lambda^{\alpha}$ dependence and depolarization by fitting our model to the single-pulse data collected on MJDs 58468 and 58469.
The recovered distributions of these two model parameters are shown for five pulse phase ranges in Extended Data Fig.~\ref{fig:single_dist}.
We expected the resulting measurements of $\alpha$ would be narrowly distributed about some preferred wavelength dependence while the recovered GRMs follow a broader distribution, reflective of changes in the underlying plasma density.
However, there is instead a strong peak in the GRM at values between 2--5\,rad\,m$^{-\alpha}$ while $\alpha$ measurements are distributed along a broad range of values that resemble the time-averaged results for these two epochs.
As a result, we can rule out rapid temporal variations in the intervening medium density as causing the low measured values of $\alpha$ and model inconsistencies.
We also looked at whether the presence of multiple over-dense regions (Faraday screens) could explain the deviations by comparing the single-pulse fits to four sets of simulated polarization spectra that were generated using our model with a fixed $\alpha = 1$ (see Methods and Extended Data Table~\ref{tab:sim_priors}).
Our recovered GRM and $\alpha$ for these simulations are shown in Extended Data Fig.~\ref{fig:sim_fits}.
All four panels display varying degrees of clustering around the true $\alpha$, along with a clear correlation between the recovered GRM and wavelength scaling. 
The pile-up of measurements where $\alpha \rightarrow 0$ and the GRM rapidly increases in addition to the long tail in the $\alpha$ distributions are caused by the simulated spectra deviating from what our model could accurately reproduce.
Simulations where the injected GRM were drawn from a Gaussian (panels b and d in Extended Data Fig.~\ref{fig:sim_fits}) provide the closest match to the observed single-pulse measurements, however the substantial peak around a particular wavelength dependence is not present in our observations of XTE~J1810$-$197.
The lack of a peak in $\alpha$ among the real data could be due to conversion within individual Faraday screens arising from different physical processes, and therefore imparting different wavelength dependencies. 
However this same behavior could very well arise from a different type of birefringent propagation, such as mode coupling, without needing to invoke more complex averaging over temporal variations or multiple Faraday screens.

\begin{table}
\centering
    \caption{Priors for the Faraday conversion model.}
    \begin{tabular}{lll}
         Parameter &  Prior distribution & Units \\ 
         \hline
         GRM                    & $\textrm{Uniform}(0, 100)$         & rad\,m$^{-\alpha}$   \\
         $\alpha$               & $\textrm{Uniform}(0, 5)$           & -                     \\
         $\Psi_{0}$             & $\textrm{Uniform}(-90, 90)$        & deg                   \\
         $\chi_{0}$             & $\textrm{Uniform}(-45, 45)$        & deg                   \\
         $\varphi$              & $\textrm{Uniform}(0, 360)$         & deg                   \\
         $\vartheta$            & $\textrm{Uniform}(0, 180)$         & deg                   \\
         $\sigma_{\rm F}$       & $\textrm{LogUniform}(10^{-4}, 10)$ & -                     \\
         $\sigma_{\rm Q}$       & $\textrm{LogUniform}(10^{-4}, 1)$  & -                     \\
         \hline
    \end{tabular}
    \label{tab:gfr_priors}
\end{table}

\section*{Discussion}

\subsection*{Interpretation of the polarization properties}

Our observations of XTE~J1810$-$197 and characterization of the variable polarization properties are challenging to interpret.
While it is possible that the observed behavior could be due to an intrinsic emission process (see ref.\cite{Philippov2022} and references therein), we have ruled out intrinsic orthogonal emission modes as a means for generating the rapid changes in PA and circular polarization across the central component.
The single-pulse PA and EA distributions across the central profile component in Fig.~\ref{fig:pa_chi} deviate by only 45\,deg from the dominant emission mode. 
Similar behavior has been described in other pulsars and magnetars as arising from some form of propagation effect in the near-field regime\cite{Edwards2004, Lower2021a}.
Exactly what kind of propagation effect is taking place depends on the environment the outgoing radio waves propagate through, and whether the wave modes of the emitted electromagnetic radiation couple to the intervening medium or themselves.

The rotation of the polarization data about an arbitrary axis in Fig.~\ref{fig:181219_bin534} (and the animated versions in the Supplementary Materials) visually appeared consistent with the radio waves from the magnetar having undergone Faraday conversion\cite{Sazonov1969}.
Unlike the propagation through the magnetized interstellar medium (which we pre-corrected for in these data), Faraday conversion results in a rotation of the polarization vector that is not restricted to the $Q$-$U$ plane.
Under this interpretation, the values of $\alpha$ and $\vartheta$ can be directly related to the physical process driving the conversion.
A relativistic pair-plasma threaded by a uniform magnetic field perpendicular to the wave direction will have linear polarization modes ($\vartheta = 90$\,deg) and impart a linear-to-circular conversion with an $\alpha = 3$ dependence\cite{Kennett1998}.
Propagation through the near-wind of a magnetar can result in values of $\alpha$ that range between 1--2 depending on the size of the phase shift that is applied to the propagating wave modes\cite{Lyutikov2022}.
Our model fits in Fig.~\ref{fig:gfr_fits} show the polarization data are consistent with propagation through a plasma with highly variable, elliptical modes with a small region of pulse phase between 0.495--0.507 being consistent with purely linear modes.
Yet the recovered values of $\alpha$ range between 0-1 with some level of clustering around $0.5$, inconsistent with theoretical expectations. 
These models also predict the fraction of polarization should remain constant with frequency, which is decidedly not the case for XTE~J1810$-$197.
Hence Faraday conversion alone cannot fully describe the polarization variability that we detected.
Other effects, such as charge imbalances, differences in electron/positron streaming velocities and the number-density ratio of cold versus and relativistic particles in the dispersive plasma can all give rise to the elliptically polarized modes that we detect\cite{Pacholczyk1973, Allen1982, Kennett1998, Luo2002}.
Changes in the magnetic field direction and mixtures of plasma modes arising from inhomogeneities in charged particle density or magnetic field strength can result in the observed damping of polarization with frequency\cite{Kennett1998}.
These phenomena are not captured by comparatively simple models of Faraday conversion.

Linear to circular conversion can also arise from a phenomenon known as mode-coupling. 
Here, differences in refractive index introduce a delay in one of the linearly polarized modes that coherently recombine at the polarization limiting radius (see ref.\cite{Petrova2001}).
This results in a fraction of the initially linearly polarized radio waves being converted to circular\cite{Melrose1977}.
Mode-coupling can occur when the magnetic field in a pair plasma that was initially quasi-tangential to the wave direction suddenly diverges from the ray path.
The change in field direction can result from bending of the field lines due to the rotation of the star\cite{Wang2010} or the presence of a twist in the magnetic field \cite{Melrose1995}.
Numerical ray tracing simulations of radio waves propagating through highly asymmetric electron-ion plasma that included this effect resulted in the transmitted position angle displaying wavelength dependencies of $\lambda^{0.5}$, as opposed to the usual quadratic dependence of standard Faraday rotation\cite{Wang2010}. 
Partially coherent addition of two modes with differing spectral indices or a frequency-dependent coherence fraction could explain both the intense polarization variations and depolarization that we observed in XTE~J1810$-$197 (see refs.\cite{Karastergiou2005, Oswald2023}).
It would also explain the unusual non-orthogonal deviations in EA and PA in Fig.~\ref{fig:pa_chi}. 
The increasing circular polarization and unusually small wavelength dependence in the later observations could be explained by changes in the dominant polarization mode over time.
These effects could be further tested by extending the recently developed partial coherence model of ref.\cite{Oswald2023} into the frequency domain.
Observations of future outbursts in XTE J1810$-$197 at frequencies below 1\, GHz would also provide strong constraints on the presence of either wave-mode coupling or Faraday conversion through detections of complete depolarization or the presence multiple oscillatory cycles in Stokes $Q$, $U$ and $V$.

The emission from XTE J1810$-$197 becomes depolarized at the same pulse phases where the total intensity takes on an inverted spectral shape.
Such behavior may be the result of synchrotron absorption in the intervening magnetospheric plasma, where the frequency dependence of synchrotron absorption can alter the spectral shape and radio polarization characteristics\cite{Luo2001, Li2023}.
Synchrotron absorption is a maximum at a frequency determined by the energy spectrum of the absorbing particles. 
A decrease in the energy of these particles causes the maximum in the absorption coefficient to move to lower frequencies, and hence the turnover in the spectrum to move to lower frequencies.
This would explain a purported shift in the spectral turnover of XTE~J1810$-$197 from higher to lower frequencies\cite{Maan2022}.
We note, however that profile components with differing spectral indices are commonplace among pulsars, and that magnetar spectral indices have been observed to undergo large fluctuations\cite{Majid2017, Lower2021a}.

\subsection*{Links to changing magnetic topology and precession}

The onset of the frequency-dependent polarization variations in XTE~J1810$-$197 occurred following an inversion of the linear PA swing.
Similar PA-swing inversions were reported in observations of both PSR J1119$-$6127 and Swift J1818.0$-$1607 during their respective 2016 and 2020 outbursts\cite{Dai2018, Lower2021a} and these were speculated to stem from a combination of changes in magnetic-field topology and birefringent effects.
Such behavior can be triggered by plastic motion of a crustal plate which would twist and distort the active magnetic field lines or even shift the location of the magnetic pole on the neutron star surface (cf. SGR~1830$-$0645\cite{Younes2022a}).
Fits to the X-ray spectrum of XTE~J1810$-$197 are consistent with the thermal X-ray emission being generated by two closely spaced, non-concentric hotspots on the surface surrounded by a warm cap\cite{Borghese2021}. 
However, the lack of high-cadence X-ray observations means it is unknown whether the surface hotspots displayed plastic motion at the same time as the polarization variations presented here.
Conversely, changes in emission height and underlying position angle swing across the pulse profile can also arise from precession of the neutron star due to either spin-orbit coupling with a binary companion (cf. PSR~1906$+$0746\cite{Desvignes2019}) or a deviation of the neutron star crust away from spheroidal symmetry resulting in free-precession\cite{Pines1974}.
Spin-orbit precession would require XTE~J1810$-$197 to have a massive companion, for which no such evidence exists.
Free-precession has been explored as a plausible explanation for the secular PA-swing evolution of XTE~J1810$-$197 following the 2018 outburst, in which the PA inversions are due to the magnetic axis of the star crossing our line of sight\cite{Desvignes2023}.
Radiation emitted close to the magnetic axis will originate from closer to the neutron star surface\cite{Melrose2014}, increasing the chance of the radio waves being subject to birefringent propagation effects from interactions with intense particle outflows in the post-outburst magnetosphere\cite{Harding1999, Tong2013}.
Mode-coupling is also predicted to be strongest for lines of sight that pass nearest to the magnetic axis, where our line of sight cuts deeper into the twisted magnetosphere\cite{Wang2010}.
Both scenarios can explain the rapid increase in circular polarization in Fig.~\ref{fig:stokes_time} just prior to the PA-swing inversion and the rapid decay afterwards.
The variations in linear-to-circular conversion with pulse-phase may then be due to local twists in the magnetic field lines or fluctuations in particle outflow along different lines of sight through the magnetosphere.


\section*{Implications}

\subsection*{Parallels with rotation-powered pulsars}

Studies of rotation-powered pulsars have shown apparent rotation measure (RM) variations as a function of pulse phase, independent of the RM induced by the magnetized interstellar medium (ISM)\cite{Ramachandran2004, Noutsos2009, Dai2015, Ilie2019}.
Some of these variations in the `phase-resolved' RM can be ascribed to scatter broadening\cite{Karastergiou2009}, but a magnetospheric or near-field propagation effect such as Faraday conversion provides the most plausible explanation.
As a demonstration of how linear-to-circular conversion can give rise to this effect, we applied the phase-resolved RM fitting approach to our observations of XTE J1810$-$197, the results of which are presented in Extended Data Fig.~\ref{fig:rm_fits}.
Significant positive and negative deviations away from the nominal ISM-induced RM of $74.4$\,rad\,m$^{-2}$ are measured across the central profile component that match-up with the rotation-phases where we detected significant conversion (see Fig.~\ref{fig:gfr_fits}).
Hence, care must be taken when interpreting apparent differential RMs across radio profiles.

\subsection*{Links to birefringent effects in fast radio bursts}

Magnetars are often invoked as the central engine of the fast radio burst (FRB) phenomenon. 
This concept was seemingly confirmed after the detection of an FRB and several other radio bursts from the Galactic magnetar SGR 1935$+$2154\cite{CHIME2020, STARE2020}.
Parallels have also been drawn between the secular RM evolution found in two repeating FRB sources\cite{Hilmarsson2021, Mckinven2022} and the Galactic Centre magnetar SGR 1745$-$2900\cite{Desvignes2018}.
The RM evolution of these objects are likely the result of slow changes in the particle density and magnetic fields of their local environments. 
More recently, large RM variations and significant levels of circular polarization have been detected in at least three other repeating FRB sources\cite{Xu2022, Feng2022b, AnnaThomas2023}.
All three are likely young objects with complex local environments as indicated by their frequency-dependent depolarization\cite{Feng2022a}, associations with active star-forming regions, persistent radio sources or both.
The RM variations in these sources may be due to transient magnetospheric propagation effects, which we demonstrated can be misidentified when fit using standard Faraday rotation models.
In the case of FRB 20201124A, Faraday conversion with wavelength dependencies between $\alpha = 2$-$3$ have been reported, suggesting the presence of a relativistic plasma surrounding the progenitor object\cite{Xu2022, Kumar2023}. 
The rapid onset and subsequent fading of intense birefringent effects in FRB 20201124A is somewhat reflected in our observations of XTE J1810$-$197 (see Fig.~\ref{fig:stokes_time}) and may point to a similar magneto-ionic environment within FRB progenitors.
Future systematic studies of birefringent propagation effects in large FRB samples and additional magnetar outbursts using current and upcoming wideband receivers could provide a novel means testing radiative transfer models in some of the most extreme magneto-ionic environments in the Universe.


\begin{methods}


\subsection*{Parkes observations}

Observations of XTE~J1810$-$197 by were performed using \textit{Murriyang}, the Parkes 64-m radio telescope, with the UWL receiver system\cite{Hobbs2020} providing continuous coverage across radio frequencies between 704-4032\,MHz. 
This includes three observations originally presented in an earlier, initial study\cite{Dai2019}.
Full Stokes data were recorded using the {\sc medusa} backend in pulsar search mode\cite{Hotan2004}, with 8-bit, 128\,$\mu$s sampling across all 3328 frequency channels (each 1\,MHz in width) which were coherently dedispersed at a dispersion measure of $178$\,pc\,cm$^{-3}$. 
We used DSPSR\cite{vanStraten2011} to fold the data at the rotation period of XTE~J1810$-$197 using an iteratively updated version of the ephemeris listed in the ATNF catalogue\cite{Manchester2005}.
These folded archives had 1024 bins covering the rotation phase of the magnetar for a time resolution of approximately 5.4\,ms.
Channels affected by radio-frequency interference (RFI) were excised in a semi-automated fashion. 
A pre-determined list of persistently bad channels were flagged using the paz tool from PSRCHIVE\cite{Hotan2004, vanStraten2012}.
This was followed by manual flagging of any remaining RFI-affected channels using the paz tool. 
We polarization calibrated these data using pac tool in PSRCHIVE to apply the calibration solutions to the data.
The solutions were derived from off-source observations of a pulsed signal injected into the signal path by a linearly polarized noise diode to determine the differential gain and phase-delays between the two recorded linear polarizations, alongside a model of the polarimetric response of the UWL inferred from applying Measurement Equation modeling\cite{vanStraten2004} to November 2018 commissioning observations of PSR~J0437$-$4715 across a large range of parallactic angles.
Faraday rotation induced by the magnetized interstellar medium between us and the source was corrected by applying the nominal rotation measure of $74.4$\,rad\,m$^{-2}$ which was obtained from observations taken on MJDs 58463, 58467 and 58470.\cite{Dai2019}.
The data presented throughout this work follows the pulsar-polarization convention\cite{vanStraten2010}, where left and right handed circular polarization are represented as negative and positive values respectively.  

Noise diode observations were not taken during P885 observations between MJD~58488-58544. 
To calibrate the affected data, we made use of close-in-time noise diode scans that were taken up to several hours before or after the corresponding observations of XTE~J1810$-$197.
In principle, the telescope system should be stable over many hours (possibly even days) with minimal instrumental distortions or polarization leakage being introduced to these data. 
To verify this, we fit the rotation measure of the magnetar at each epoch using direct fits to Stokes $Q$ and $U$ spectra extracted from the precursor component using RMNest\cite{Lower2022ascl}.
Note these measurements were not corrected for propagation through the ionosphere of the Earth, which can contribute between $-0.2$ to $-2$\,rad\,m$^{-2}$ at the location of Parkes\cite{Han2018}, which can account for small epoch-to-epoch deviations in the measured RM.

Two fold-mode observations performed on 15 and 18 December 2017 were flux calibrated using on- and off-source observations of the noise diode while pointed at or one-degree offset from the radio galaxy 3C\,218 (Hydra A) to measure the apparent brightness of the diode and correct for the system absolute gain and phase.
However, the absolute gain of early observations could not be calibrated using standard methods.
For these observations we estimate the flux density using a similar technique to that used by Scholz et al. to flux calibrate observations of PSR~J1622$-$1950\cite{Scholz2017}. 
We estimated the system equivalent flux density (SEFD) for the individual Stokes parameters as a function of frequency from the expected baseline root-mean-square (RMS) of the calibrated fold-mode observation taken on December 18 via the radiometer equation\cite{Dewey1985}
\begin{equation}\label{eqn:sefd}
    \sigma_{\mathrm{off}} = C_{\ell} \frac{\mathrm{SEFD}}{\sqrt{n_{p} \Delta\nu \Delta t}}.
\end{equation}
Here, $C_{\ell}$ is the loss factor from signal digitization, $n_{p}$ is the number of summed polarizations, $\Delta\nu$ the bandwidth and $\Delta t = T_{\mathrm{int}/n_{\mathrm{bin}}}$ is the integration time per phase bin. 
We then converted the individual Stokes parameters for the affected observation to units of Jansky by scaling the off-pulse RMS by the right-hand side of Equation~\ref{eqn:sefd} (with $n_{p} = 1$).

\bigbreak

\subsection*{Jodrell Bank observations}

XTE J1810–197 is regularly monitored with the 76-m Lovell telescope at JBO with the L-band receiver.
This is a hybrid system which detects radio waves with linear probes after passing through a quarter wavelength plate.
The length of each observation varied from 20 minutes to several hours, but most of the observations were between 30-60 minutes long. 
The radio frequency band was centered at 1,532 MHz, with a bandwidth of 384 MHz after RFI excision\cite{Caleb2022}.
The observations were folded and dedispersed online. 
The folding requires an initial timing ephemeris, which was precise enough such that the time smearing is insignificant within each 20-second long subintegration. 
Full Stokes data were recorded with 1,024 phase bins across the pulse period and 768 frequency channels.
After manual inspection and RFI excision, the data was polarization calibrated. 

The polarization response of the system was tested by monitoring the polarization of three sources: PSRs B0540+23, B0611+22 and B1737$-$30, which were observed during the same sessions.
These observations were compared to the known polarization properties of these sources by examining the differences with the published polarization profiles (see ref.\cite{Johnston2017}). 
For each observation, a frequency resolved receiver model was derived by fitting for a differential phase and gain.
This solution minimizes the frequency dependence of the polarization properties in the observations (apart from those expected from Faraday rotation), and maximizes the similarity between the observed and known polarization properties.
Given the obtained receiver solutions are similar, a smooth polynomial as function of frequency was used to describe frequency dependence of the parameters. 
These polynomials are fit to the combined parameters obtained from the individual solutions.
This single combined receiver model was applied to the XTE J1810–197 data to remove the undesired frequency dependent polarization response of the receiver.

\bigbreak
\subsection*{Phase resolved spectral indices}

We tested whether the profile components of XTE~J1810$-$197 exhibit different spectral indices by spectral fits to the total intensity spectra extracted from individual phase-bins across the profile. 
This was performed using copies of the data that were binned such that 256\,phase bins cover the rotation period of the magnetar.
We fit the data using a simple power-law function of the form
\begin{equation}
    S(x) = A\,x^{\kappa},
\end{equation}
where $x = \nu/1\,{\rm GHz}$ is the observation frequency normalized to 1\,GHz, $A$ is a linear scaling parameter and $\kappa$ the spectral index.
We used a Gaussian likelihood of the form
\begin{equation}\label{eqn:gauss_like}
	\Like(\mathbf{d} | \theta) = \prod \frac{1}{\sqrt{2\pi\sigma^{2}}} \exp \Big( -\frac{(d - \mu(\theta))^{2}}{2\sigma^{2}} \Big),
\end{equation}
in which $d$ is the measured flux density, $\theta$ the model parameters, $\sigma$ the flux density uncertainties and $\mu = S(x)$ our power-law model.
Throughout this work the posterior distributions for our various model fits to the data were sampled using Bilby\cite{bilby} as a front-end to the Dynesty nested sampling algorithm\cite{dynesty}.

\bigbreak

\subsection*{Modelling the linear-to-circular conversion}

Electromagnetic radiation that propagates through a birefringent medium with elliptical or linearly polarized natural wave modes will undergo a generalized form of Faraday rotation, known as Faraday conversion.
Unlike standard Faraday rotation, induced by intervening plasma with circularly polarized modes such as the magnetized interstellar medium, there is no simple, catchall model for describing the observational effects of Faraday conversion. 
This largely comes from the variety of environments in which Faraday conversion can be produced (see for example refs.\cite{Kennett1998, Melrose2010, Gruzinov2019, Lyutikov2022}).
Other propagation effects such as coherent/partially-coherent mode mixing can also lead to a birefringent conversion of linear to circularly polarized radio waves without a well-defined frequency dependence\cite{Melrose1977, Petrova2001, Karastergiou2005, Dyks2019, Oswald2023}
In order to model the linear-to-circular conversion detected in the radio pulses from XTE~J1810$-$197, we used the phenomenological framework described in ref.\cite{Lower2021b} which we outline as follows.

The polarization vector normalized by the total polarization, $P$, is defined as
\begin{equation}\label{eqn:poln_wave}
    \mathbf{P}_{0}(\lambda) = 
    \frac{1}{P}
    \begin{bmatrix}
    Q\\
    U\\
    V
    \end{bmatrix}
    =
    \begin{bmatrix}
    \cos[2\Psi(\lambda)]\cos(2\chi_{0})\\
    \sin[2\Psi(\lambda)]\cos(2\chi_{0})\\
    \sin(2\chi_{0})
    \end{bmatrix},
\end{equation}
where $\chi_{0}$ is the ellipticity angle and $\Psi(\lambda)$ is the linear polarization position angle as a function of wavelength.
$\Psi(\lambda)$ is expressed as
\begin{equation}\label{eqn:grm}
    \Psi(\lambda) = \Psi_{0} + {\rm GRM} (\lambda^{\alpha} - \lambda_{c}^{\alpha}),
\end{equation}
in which $\Psi_{0}$ is the reference position angle at the central observing frequency ($\lambda_{c} = 2368$\,MHz), GRM is the `generalized' rotation measure and $\alpha$ is the spectral dependence.
The linear-to-circular conversion is then emulated by applying a pair of rotation matrices to shift the longitude and latitude of the polarization-rotation axis
\begin{equation}
    \mathbf{R}_{\vartheta} = 
    \begin{bmatrix}
    \cos(\vartheta) & 0 & \sin(\vartheta)\\
    0 & 1 & 0\\
    -\sin(\vartheta) & 0 & \cos(\vartheta)
    \end{bmatrix},
\end{equation}
and
\begin{equation}\label{eqn:v_rot}
    \mathbf{R}_{\varphi} = 
    \begin{bmatrix}
    \cos(\varphi) & -\sin(\varphi) & 0\\
    \sin(\varphi) & \cos(\varphi) & 0\\
    0 & 0 & 1
    \end{bmatrix}.
\end{equation}
Here, $\vartheta$ is the angle about which the axis has been rotated about the Stokes $U$ axis (away from the positive Stokes $V$ axis) and $\varphi$ is the angular rotation applied about the Stokes $V$ axis.
The full phenomenological model can be written as
\begin{equation}\label{eqn:pol_model}
    \hat{\mathbf{P}}_{m}(\lambda)
    =
    \mathbf{R}_{\vartheta} \cdot \mathbf{R}_{\varphi} \cdot \mathbf{P}_{0}(\lambda).
\end{equation} 
In principle, the value of $\vartheta$ encodes information about the polarization modes of the intervening medium. 
For $\vartheta = 0^{\circ}$ the modes are circular, $0^{\circ} < \vartheta < 90^{\circ}$ or $90^{\circ} < \vartheta < 180^{\circ}$ the modes are elliptical, and $\vartheta = 90^{\circ}$ corresponds to purely linear modes.
In this context, `standard' Faraday rotation would correspond to $\vartheta = \varphi = 0^{\circ}$ with spectral dependence $\alpha = 2$. 

We applied the phenomenological model to the normalized Stokes $Q$, $U$ and $V$ spectra extracted from all pulse-phase bins between phases $0.490$ to $0.535$ for the time-averaged data. 
Posterior distributions for each of the model parameters were inferred using Bayesian parameter estimation. 
We used a joint Gaussian likelihood function of the form
\begin{equation}
	\Like(\mathbf{P}(\lambda) | \theta) = \prod_{i=1}^{3} \prod_{j=1}^{N} \frac{1}{\sqrt{2\pi\hat{\sigma}_{ij}^{2}}} \exp \Big[ -\frac{(\mathbf{P}_{i}(\lambda_{j}) - \hat{\mathbf{P}}_{m}^{i}(\theta;\lambda_{j}))^{2}}{2\hat{\sigma}_{ij}^{2}} \Big],
\end{equation}
where $\hat{\sigma}_{ij}^{2}$ is RMS uncertainty on the $i$-th Stokes parameter in the $j$-th frequency channel (up to $N$-frequency channels) after applying extra error scale factor (EFAC; $\sigma_{\rm F}$) and error in quadrature (EQUAD; $\sigma_{\rm Q}$) terms to correct for unaccounted systematic uncertainties as
\begin{equation}
    \hat{\sigma}_{ij}^{2} = (\sigma_{\rm F}\times \sigma_{ij})^{2} + \sigma_{\rm Q}^{2}.
\end{equation}
Parameter estimation was conducted using the RMNest package\cite{Lower2022ascl}.
The data were Faraday de-rotated at the nominal RM of $74.4$\,rad\,m$^{-2}$ and binned to a frequency resolution of 8\,MHz per channel prior to fitting the Faraday conversion model.

We initially suggested the unusually small wavelength dependencies ($\alpha \lesssim 1$) that are predominately seen between phases $0.49$ and $0.53$ may have been the result of averaging over pulse-to-pulse changes in the spectra due to stochastic plasma variations along the line of sight.
This was tested by extracting the Stokes spectra from individual single pulses from the Parkes observations taken on MJDs MJD 58468 and 58469.
Although search-mode data were also collected on MJD 58471, the resulting spectra were heavily contaminated by broadband, impulsive RFI.
We therefore chose to not include the data from this epoch in our single-pulse investigation.
Following data calibration and RFI excision, we fit the phenomenological model to a set of 6,537 Stokes spectra extracted from the single-pulses where the total intensity flux as a minimum of $\times10$ the off-pulse root-mean square value.
Radio frequencies between 704 to 1300 MHz are strongly affected by impulsive, broadband RFI, which becomes smeared across pulse phase after dedispersing the data.
To avoid both the S/N and RFI-contamination issues, we ignored all data below 1300\,MHz when conducting the single-pulse parameter estimation. 
Spectra where the posterior distributions for $\{{\rm GRM},\alpha\} \rightarrow \{0,0\}$ (i.e no significant support for Faraday conversion) at the 95\% confidence interval were not included in our final sample.
We also ignored spectra where the GRM posterior distributions were not well constrained by imposing a limit where the range of inferred GRM values encompassed by either the upper or lower confidence intervals had to be less than $\pm 20$\,rad\,m$^{-\alpha}$. 
After imposing these thresholds, we were left with 2,142 spectra for which we had confidently detected and subsequently characterized the Faraday conversion.

\bigbreak

\subsection*{Simulating multiple Faraday screens}
Our fits to the polarization spectra extracted from the Parkes-UWL single-pulse data returned a spread of wavelength dependencies that are approximately centered on $\lambda^{1}$.
This behavior was interpreted as potentially resulting from one of two, potentially related scenarios: the relatively simple particle wind model fails to fully capture the complex electrodynamics that take place within the magnetosphere/near-wind environment of the magnetar, or there are more than one over-dense `Faraday screens' in which the radiation has propagated through. 
Under the latter hypothesis, the broad distribution of wavelength dependencies inferred from our single-pulse fits would arise from incoherent addition of multiple rotations of the polarization vector along the line of sight. 

We performed four sets of simulations using the phenomenological Faraday conversion model to emulate this behavior, two of which were designed to test the possible impact of averaging over two orthogonally polarized modes. 
These `OPM' simulations used Gaussian priors on $\varphi$, the means of which were separated by 180\,degrees.
The other two simulations sampled $\varphi$ Uniformly between 0-180\,degrees.
We also tested the impact of drawing the injected GRM values from either a Uniform or Gaussian distribution on the resulting scatter in the recovered $\alpha$. 
After drawing the initial parameters, we generated two individual Stokes spectra in each simulation, both with the wavelength dependence held fixed at $\alpha = 1$.
An `observed' spectrum was then created by incoherently summing these two initial spectra, which we then injected into Gaussian noise.
Our measurement uncertainties for the simulated Stokes spectra were drawn from the median Stokes parameter uncertainties measured on MJD 59871.
We then applied the same parameter estimation framework that was used for characterizing the real observations. 

\bigbreak

\subsection*{Phase-resolved Faraday rotation}

Phase-resolved studies of rotation powered pulsars have discovered apparent, rotation-phase dependent deviations in their RM of away from the nominal ISM-induced values\cite{Ramachandran2004, Noutsos2009, Dai2015, Ilie2019}.
These deviations have been speculated to arise from propagation effects within the magnetospheres of these pulsars.
Several repeating FRBs have also been found to exhibit transient, rapid RM variations over relatively short (month to year) time spans\cite{Xu2022, AnnaThomas2023}, with peak $\Delta$RMs of up to $\pm$36,000\,rad\,m$^{-2}$.
Intense Faraday conversion has been identified in least one one of these sources to date, which if not accounted for may be confused for large RM variations\cite{Xu2022, Kumar2023}.
To demonstrate the impact of unmodelled Faraday conversion on measurements of `standard' Faraday rotation, we performed a set of phase-resolved rotation measure fits to the MJD 58463-58471 observations of XTE~J1810$-$197.
We utilized the same RMNest-based approach used for modeling the Faraday conversion, but with a fixed wavelength dependence of $\alpha = 2$ and the polarization rotation axis longitude and latitude held fixed at the $V$-axis pole (i.e $\{ \varphi, \vartheta \} = \{ 0, 0 \}$).

\end{methods}

\bigbreak

\textbf{Data availability}
Raw Parkes data files (totaling 3.1\,TB) are available to download from the CSIRO Data Access Portal (\url{https://data.csiro.au/}).
The time and frequency averaged polarization profiles collected by the Lovell Telescope are available on Zenodo: \url{https://doi.org/10.5281/zenodo.10595269}.

\textbf{Code availability}
Fold-mode data products were produced from the raw Parkes search-mode data using the DSPSR\cite{vanStraten2011}, which were then calibrated, cleaned and processed using the software tools in the PSRCHIVE\cite{Hotan2004, vanStraten2012} package.
PSRCHIVE was also used for processing the folded Lovell Telescope data.
Measurements of the linear-to-circular conversion phenomena was performed using and analyzed using RMNest\cite{Lower2022ascl}  Python package, which used Bilby\cite{bilby} and dynesty\cite{dynesty} to sample the model parameters. 
Specific Python scripts used in this analysis are available on request from M.E.L.


\bibliography{main}

\begin{thebibliography}{10}
\expandafter\ifx\csname url\endcsname\relax
  \def\url#1{\texttt{#1}}\fi
\expandafter\ifx\csname urlprefix\endcsname\relax\def\urlprefix{URL }\fi
\providecommand{\bibinfo}[2]{#2}
\providecommand{\eprint}[2][]{\url{#2}}

\bibitem{Olausen2014}
\bibinfo{author}{{Olausen}, S.~A.} \& \bibinfo{author}{{Kaspi}, V.~M.}
\newblock \bibinfo{title}{{The McGill Magnetar Catalog}}.
\newblock \emph{\bibinfo{journal}{\apjs}} \textbf{\bibinfo{volume}{212}},
  \bibinfo{pages}{6} (\bibinfo{year}{2014}).
\newblock \eprint{1309.4167}.

\bibitem{Ibrahim2004}
\bibinfo{author}{{Ibrahim}, A.~I.} \emph{et~al.}
\newblock \bibinfo{title}{{Discovery of a Transient Magnetar: XTE J1810-197}}.
\newblock \emph{\bibinfo{journal}{\apjl}} \textbf{\bibinfo{volume}{609}},
  \bibinfo{pages}{L21--L24} (\bibinfo{year}{2004}).
\newblock \eprint{astro-ph/0310665}.

\bibitem{Halpern2005}
\bibinfo{author}{{Halpern}, J.~P.}, \bibinfo{author}{{Gotthelf}, E.~V.},
  \bibinfo{author}{{Becker}, R.~H.}, \bibinfo{author}{{Helfand }, D.~J.} \&
  \bibinfo{author}{{White}, R.~L.}
\newblock \bibinfo{title}{{Discovery of Radio Emission from the Transient
  Anomalous X-Ray Pulsar XTE J1810-197}}.
\newblock \emph{\bibinfo{journal}{\apjl}} \textbf{\bibinfo{volume}{632}},
  \bibinfo{pages}{L29--L32} (\bibinfo{year}{2005}).
\newblock \eprint{astro-ph/0508534}.

\bibitem{Camilo2006}
\bibinfo{author}{{Camilo}, F.} \emph{et~al.}
\newblock \bibinfo{title}{{Transient pulsed radio emission from a magnetar}}.
\newblock \emph{\bibinfo{journal}{\nat}} \textbf{\bibinfo{volume}{442}},
  \bibinfo{pages}{892--895} (\bibinfo{year}{2006}).
\newblock \eprint{astro-ph/0605429}.

\bibitem{Camilo2007b}
\bibinfo{author}{{Camilo}, F.} \emph{et~al.}
\newblock \bibinfo{title}{{The Variable Radio-to-X-Ray Spectrum of the Magnetar
  XTE J1810-197}}.
\newblock \emph{\bibinfo{journal}{\apj}} \textbf{\bibinfo{volume}{669}},
  \bibinfo{pages}{561--569} (\bibinfo{year}{2007}).
\newblock \eprint{0705.4095}.

\bibitem{Kramer2007}
\bibinfo{author}{{Kramer}, M.}, \bibinfo{author}{{Stappers}, B.~W.},
  \bibinfo{author}{{Jessner}, A.}, \bibinfo{author}{{Lyne}, A.~G.} \&
  \bibinfo{author}{{Jordan}, C.~A.}
\newblock \bibinfo{title}{{Polarized radio emission from a magnetar}}.
\newblock \emph{\bibinfo{journal}{\mnras}} \textbf{\bibinfo{volume}{377}},
  \bibinfo{pages}{107--119} (\bibinfo{year}{2007}).
\newblock \eprint{astro-ph/0702365}.

\bibitem{Goldreich1969}
\bibinfo{author}{{Goldreich}, P.} \& \bibinfo{author}{{Julian}, W.~H.}
\newblock \bibinfo{title}{{Pulsar Electrodynamics}}.
\newblock \emph{\bibinfo{journal}{\apj}} \textbf{\bibinfo{volume}{157}},
  \bibinfo{pages}{869} (\bibinfo{year}{1969}).

\bibitem{Sazonov1969}
\bibinfo{author}{{Sazonov}, V.~N.}
\newblock \bibinfo{title}{{Generation and Transfer of Polarized Synchrotron
  Radiation.}}
\newblock \emph{\bibinfo{journal}{Soviet Astron.}}
  \textbf{\bibinfo{volume}{13}}, \bibinfo{pages}{396} (\bibinfo{year}{1969}).

\bibitem{Kennett1998}
\bibinfo{author}{{Kennett}, M.} \& \bibinfo{author}{{Melrose}, D.}
\newblock \bibinfo{title}{{Propagation-induced Circular Polarisation in
  Synchrotron Sources}}.
\newblock \emph{\bibinfo{journal}{\pasa}} \textbf{\bibinfo{volume}{15}},
  \bibinfo{pages}{211--216} (\bibinfo{year}{1998}).

\bibitem{Edwards2004}
\bibinfo{author}{{Edwards}, R.~T.} \& \bibinfo{author}{{Stappers}, B.~W.}
\newblock \bibinfo{title}{{Ellipticity and deviations from orthogonality in the
  polarization modes of PSR B0329+54}}.
\newblock \emph{\bibinfo{journal}{\aap}} \textbf{\bibinfo{volume}{421}},
  \bibinfo{pages}{681--691} (\bibinfo{year}{2004}).
\newblock \eprint{astro-ph/0404092}.

\bibitem{Noutsos2009}
\bibinfo{author}{{Noutsos}, A.}, \bibinfo{author}{{Karastergiou}, A.},
  \bibinfo{author}{{Kramer}, M.}, \bibinfo{author}{{Johnston}, S.} \&
  \bibinfo{author}{{Stappers}, B.~W.}
\newblock \bibinfo{title}{{Phase-resolved Faraday rotation in pulsars}}.
\newblock \emph{\bibinfo{journal}{\mnras}} \textbf{\bibinfo{volume}{396}},
  \bibinfo{pages}{1559--1572} (\bibinfo{year}{2009}).
\newblock \eprint{0903.5511}.

\bibitem{Ilie2019}
\bibinfo{author}{{Ilie}, C.~D.}, \bibinfo{author}{{Johnston}, S.} \&
  \bibinfo{author}{{Weltevrede}, P.}
\newblock \bibinfo{title}{{Evidence for magnetospheric effects on the radiation
  of radio pulsars}}.
\newblock \emph{\bibinfo{journal}{\mnras}} \textbf{\bibinfo{volume}{483}},
  \bibinfo{pages}{2778--2794} (\bibinfo{year}{2019}).
\newblock \eprint{1811.12831}.

\bibitem{Sobey2021}
\bibinfo{author}{{Sobey}, C.} \emph{et~al.}
\newblock \bibinfo{title}{{A polarization census of bright pulsars using the
  ultrawideband receiver on the Parkes radio telescope}}.
\newblock \emph{\bibinfo{journal}{\mnras}} \textbf{\bibinfo{volume}{504}},
  \bibinfo{pages}{228--247} (\bibinfo{year}{2021}).
\newblock \eprint{2103.13838}.

\bibitem{Li2023}
\bibinfo{author}{{Li}, D.}, \bibinfo{author}{{Bilous}, A.},
  \bibinfo{author}{{Ransom}, S.}, \bibinfo{author}{{Main}, R.} \&
  \bibinfo{author}{{Yang}, Y.-P.}
\newblock \bibinfo{title}{{A highly magnetized environment in a pulsar binary
  system}}.
\newblock \emph{\bibinfo{journal}{\nat}} \textbf{\bibinfo{volume}{618}},
  \bibinfo{pages}{484--488} (\bibinfo{year}{2023}).
\newblock \eprint{2205.07917}.

\bibitem{Melrose1977}
\bibinfo{author}{{Melrose}, D.~B.} \& \bibinfo{author}{{Stoneham}, R.~J.}
\newblock \bibinfo{title}{{The natural wave modes in a pulsar magnetosphere.}}
\newblock \emph{\bibinfo{journal}{\pasa}} \textbf{\bibinfo{volume}{3}},
  \bibinfo{pages}{120--122} (\bibinfo{year}{1977}).

\bibitem{Petrova2001}
\bibinfo{author}{{Petrova}, S.~A.}
\newblock \bibinfo{title}{{On the origin of orthogonal polarization modes in
  pulsar radio emission}}.
\newblock \emph{\bibinfo{journal}{\aap}} \textbf{\bibinfo{volume}{378}},
  \bibinfo{pages}{883--897} (\bibinfo{year}{2001}).

\bibitem{Dyks2019}
\bibinfo{author}{{Dyks}, J.}
\newblock \bibinfo{title}{{Radio pulsar polarization as a coherent sum of
  orthogonal proper mode waves}}.
\newblock \emph{\bibinfo{journal}{\mnras}} \textbf{\bibinfo{volume}{488}},
  \bibinfo{pages}{2018--2040} (\bibinfo{year}{2019}).
\newblock \eprint{1902.11141}.

\bibitem{Oswald2023}
\bibinfo{author}{{Oswald}, L.~S.}, \bibinfo{author}{{Karastergiou}, A.} \&
  \bibinfo{author}{{Johnston}, S.}
\newblock \bibinfo{title}{{Pulsar polarization: a partial-coherence model}}.
\newblock \emph{\bibinfo{journal}{\mnras}} \textbf{\bibinfo{volume}{525}},
  \bibinfo{pages}{840--853} (\bibinfo{year}{2023}).
\newblock \eprint{2307.14265}.

\bibitem{Levin2019}
\bibinfo{author}{{Levin}, L.} \emph{et~al.}
\newblock \bibinfo{title}{{Spin frequency evolution and pulse profile
  variations of the recently re-activated radio magnetar XTE J1810-197}}.
\newblock \emph{\bibinfo{journal}{\mnras}} \textbf{\bibinfo{volume}{488}},
  \bibinfo{pages}{5251--5258} (\bibinfo{year}{2019}).
\newblock \eprint{1903.02660}.

\bibitem{Gotthelf2019}
\bibinfo{author}{{Gotthelf}, E.~V.} \emph{et~al.}
\newblock \bibinfo{title}{{The 2018 X-Ray and Radio Outburst of Magnetar XTE
  J1810-197}}.
\newblock \emph{\bibinfo{journal}{\apjl}} \textbf{\bibinfo{volume}{874}},
  \bibinfo{pages}{L25} (\bibinfo{year}{2019}).
\newblock \eprint{1902.08358}.

\bibitem{Dai2019}
\bibinfo{author}{{Dai}, S.} \emph{et~al.}
\newblock \bibinfo{title}{{Wideband Polarized Radio Emission from the Newly
  Revived Magnetar XTE J1810-197}}.
\newblock \emph{\bibinfo{journal}{\apjl}} \textbf{\bibinfo{volume}{874}},
  \bibinfo{pages}{L14} (\bibinfo{year}{2019}).
\newblock \eprint{1902.04689}.

\bibitem{Maan2019}
\bibinfo{author}{{Maan}, Y.}, \bibinfo{author}{{Joshi}, B.~C.},
  \bibinfo{author}{{Surnis}, M.~P.}, \bibinfo{author}{{Bagchi}, M.} \&
  \bibinfo{author}{{Manoharan}, P.~K.}
\newblock \bibinfo{title}{{Distinct Properties of the Radio Burst Emission from
  the Magnetar XTE J1810-197}}.
\newblock \emph{\bibinfo{journal}{\apjl}} \textbf{\bibinfo{volume}{882}},
  \bibinfo{pages}{L9} (\bibinfo{year}{2019}).
\newblock \eprint{1908.04304}.

\bibitem{Manchester1975}
\bibinfo{author}{{Manchester}, R.~N.}, \bibinfo{author}{{Taylor}, J.~H.} \&
  \bibinfo{author}{{Huguenin}, G.~R.}
\newblock \bibinfo{title}{{Observations of pulsar radio emission. II.
  Polarization of individual pulses.}}
\newblock \emph{\bibinfo{journal}{\apj}} \textbf{\bibinfo{volume}{196}},
  \bibinfo{pages}{83--102} (\bibinfo{year}{1975}).

\bibitem{Karastergiou2005}
\bibinfo{author}{{Karastergiou}, A.}, \bibinfo{author}{{Johnston}, S.} \&
  \bibinfo{author}{{Manchester}, R.~N.}
\newblock \bibinfo{title}{{Polarization profiles of southern pulsars at 3.1
  GHz}}.
\newblock \emph{\bibinfo{journal}{\mnras}} \textbf{\bibinfo{volume}{359}},
  \bibinfo{pages}{481--492} (\bibinfo{year}{2005}).
\newblock \eprint{astro-ph/0502337}.

\bibitem{Dyks2020}
\bibinfo{author}{{Dyks}, J.}
\newblock \bibinfo{title}{{Artefacts of circumpolar cartography in radio pulsar
  polarization}}.
\newblock \emph{\bibinfo{journal}{\mnras}} \textbf{\bibinfo{volume}{495}},
  \bibinfo{pages}{L118--L122} (\bibinfo{year}{2020}).
\newblock \eprint{2005.12135}.

\bibitem{Lower2021b}
\bibinfo{author}{{Lower}, M.~E.}
\newblock \bibinfo{title}{{A phenomenological model for measuring generalised
  Faraday rotation}}.
\newblock \emph{\bibinfo{journal}{arXiv e-prints}}
  \bibinfo{pages}{arXiv:2108.09429} (\bibinfo{year}{2021}).
\newblock \eprint{2108.09429}.

\bibitem{Gruzinov2019}
\bibinfo{author}{{Gruzinov}, A.} \& \bibinfo{author}{{Levin}, Y.}
\newblock \bibinfo{title}{{Conversion Measure of Faraday Rotation-Conversion
  with Application to Fast Radio Bursts}}.
\newblock \emph{\bibinfo{journal}{\apj}} \textbf{\bibinfo{volume}{876}},
  \bibinfo{pages}{74} (\bibinfo{year}{2019}).
\newblock \eprint{1902.01485}.

\bibitem{Lyutikov2022}
\bibinfo{author}{{Lyutikov}, M.}
\newblock \bibinfo{title}{{Faraday Conversion in Pair-symmetric Winds of
  Magnetars and Fast Radio Bursts}}.
\newblock \emph{\bibinfo{journal}{\apjl}} \textbf{\bibinfo{volume}{933}},
  \bibinfo{pages}{L6} (\bibinfo{year}{2022}).
\newblock \eprint{2205.13435}.

\bibitem{Philippov2022}
\bibinfo{author}{{Philippov}, A.} \& \bibinfo{author}{{Kramer}, M.}
\newblock \bibinfo{title}{{Pulsar Magnetospheres and Their Radiation}}.
\newblock \emph{\bibinfo{journal}{\araa}} \textbf{\bibinfo{volume}{60}},
  \bibinfo{pages}{495--558} (\bibinfo{year}{2022}).

\bibitem{Lower2021a}
\bibinfo{author}{{Lower}, M.~E.}, \bibinfo{author}{{Johnston}, S.},
  \bibinfo{author}{{Shannon}, R.~M.}, \bibinfo{author}{{Bailes}, M.} \&
  \bibinfo{author}{{Camilo}, F.}
\newblock \bibinfo{title}{{The dynamic magnetosphere of Swift J1818.0-1607}}.
\newblock \emph{\bibinfo{journal}{\mnras}} \textbf{\bibinfo{volume}{502}},
  \bibinfo{pages}{127--139} (\bibinfo{year}{2021}).
\newblock \eprint{2011.12463}.

\bibitem{Pacholczyk1973}
\bibinfo{author}{{Pacholczyk}, A.~G.}
\newblock \bibinfo{title}{{Circular repolarization in compact radio sources}}.
\newblock \emph{\bibinfo{journal}{\mnras}} \textbf{\bibinfo{volume}{163}},
  \bibinfo{pages}{29P} (\bibinfo{year}{1973}).

\bibitem{Allen1982}
\bibinfo{author}{{Allen}, M.~C.} \& \bibinfo{author}{{Melrose}, D.~B.}
\newblock \bibinfo{title}{{Elliptically polarized natural modes in pulsar
  magnetospheres}}.
\newblock \emph{\bibinfo{journal}{\pasa}} \textbf{\bibinfo{volume}{4}},
  \bibinfo{pages}{365--370} (\bibinfo{year}{1982}).

\bibitem{Luo2002}
\bibinfo{author}{{Luo}, Q.}, \bibinfo{author}{{Melrose}, D.~B.} \&
  \bibinfo{author}{{Fussell}, D.}
\newblock \bibinfo{title}{{Wave dispersion in gyrotropic relativistic pulsar
  plasmas}}.
\newblock \emph{\bibinfo{journal}{\pre}} \textbf{\bibinfo{volume}{66}},
  \bibinfo{pages}{026405} (\bibinfo{year}{2002}).

\bibitem{Wang2010}
\bibinfo{author}{{Wang}, C.}, \bibinfo{author}{{Lai}, D.} \&
  \bibinfo{author}{{Han}, J.}
\newblock \bibinfo{title}{{Polarization changes of pulsars due to wave
  propagation through magnetospheres}}.
\newblock \emph{\bibinfo{journal}{\mnras}} \textbf{\bibinfo{volume}{403}},
  \bibinfo{pages}{569--588} (\bibinfo{year}{2010}).
\newblock \eprint{0910.2793}.

\bibitem{Melrose1995}
\bibinfo{author}{{Melrose}, D.~B.}, \bibinfo{author}{{Robinson}, P.~A.} \&
  \bibinfo{author}{{Feletto}, T.~M.}
\newblock \bibinfo{title}{{Mode Coupling Due to Twisting of Magnetic Field
  Lines}}.
\newblock \emph{\bibinfo{journal}{\solphys}} \textbf{\bibinfo{volume}{158}},
  \bibinfo{pages}{139--158} (\bibinfo{year}{1995}).

\bibitem{Luo2001}
\bibinfo{author}{{Luo}, Q.} \& \bibinfo{author}{{Melrose}, D.~B.}
\newblock \bibinfo{title}{{Cyclotron absorption of radio emission within pulsar
  magnetospheres}}.
\newblock \emph{\bibinfo{journal}{\mnras}} \textbf{\bibinfo{volume}{325}},
  \bibinfo{pages}{187--196} (\bibinfo{year}{2001}).

\bibitem{Maan2022}
\bibinfo{author}{{Maan}, Y.}, \bibinfo{author}{{Surnis}, M.~P.},
  \bibinfo{author}{{Chandra Joshi}, B.} \& \bibinfo{author}{{Bagchi}, M.}
\newblock \bibinfo{title}{{Magnetar XTE J1810-197: Spectro-temporal Evolution
  of Average Radio Emission}}.
\newblock \emph{\bibinfo{journal}{\apj}} \textbf{\bibinfo{volume}{931}},
  \bibinfo{pages}{67} (\bibinfo{year}{2022}).
\newblock \eprint{2201.13006}.

\bibitem{Majid2017}
\bibinfo{author}{{Majid}, W.~A.} \emph{et~al.}
\newblock \bibinfo{title}{{Post-outburst Radio Observations of the High
  Magnetic Field Pulsar PSR J1119-6127}}.
\newblock \emph{\bibinfo{journal}{\apjl}} \textbf{\bibinfo{volume}{834}},
  \bibinfo{pages}{L2} (\bibinfo{year}{2017}).
\newblock \eprint{1612.02868}.

\bibitem{Dai2018}
\bibinfo{author}{{Dai}, S.} \emph{et~al.}
\newblock \bibinfo{title}{{Peculiar spin frequency and radio profile evolution
  of PSR J1119-6127 following magnetar-like X-ray bursts}}.
\newblock \emph{\bibinfo{journal}{\mnras}} \textbf{\bibinfo{volume}{480}},
  \bibinfo{pages}{3584--3594} (\bibinfo{year}{2018}).
\newblock \eprint{1806.05064}.

\bibitem{Younes2022a}
\bibinfo{author}{{Younes}, G.} \emph{et~al.}
\newblock \bibinfo{title}{{Pulse Peak Migration during the Outburst Decay of
  the Magnetar SGR 1830-0645: Crustal Motion and Magnetospheric Untwisting}}.
\newblock \emph{\bibinfo{journal}{\apjl}} \textbf{\bibinfo{volume}{924}},
  \bibinfo{pages}{L27} (\bibinfo{year}{2022}).
\newblock \eprint{2201.05517}.

\bibitem{Borghese2021}
\bibinfo{author}{{Borghese}, A.} \emph{et~al.}
\newblock \bibinfo{title}{{The X-ray evolution and geometry of the 2018
  outburst of XTE J1810-197}}.
\newblock \emph{\bibinfo{journal}{\mnras}} \textbf{\bibinfo{volume}{504}},
  \bibinfo{pages}{5244--5257} (\bibinfo{year}{2021}).
\newblock \eprint{2104.11083}.

\bibitem{Desvignes2019}
\bibinfo{author}{{Desvignes}, G.} \emph{et~al.}
\newblock \bibinfo{title}{{Radio emission from a pulsar{\textquoteright}s
  magnetic pole revealed by general relativity}}.
\newblock \emph{\bibinfo{journal}{Science}} \textbf{\bibinfo{volume}{365}},
  \bibinfo{pages}{1013--1017} (\bibinfo{year}{2019}).
\newblock \eprint{1909.06212}.

\bibitem{Pines1974}
\bibinfo{author}{{Pines}, D.}
\newblock \bibinfo{title}{{Free precession of neutron stars}}.
\newblock \emph{\bibinfo{journal}{\nat}} \textbf{\bibinfo{volume}{248}},
  \bibinfo{pages}{483--486} (\bibinfo{year}{1974}).

\bibitem{Desvignes2023}
\bibinfo{author}{{Desvignes}, G.} \emph{et~al.}
\newblock \bibinfo{title}{{A freely precessing magnetar following an X-ray
  outburst}}  (\bibinfo{year}{Submitted}).

\bibitem{Melrose2014}
\bibinfo{author}{{Melrose}, D.~B.} \& \bibinfo{author}{{Yuen}, R.}
\newblock \bibinfo{title}{{Non-corotating models for pulsar magnetospheres}}.
\newblock \emph{\bibinfo{journal}{\mnras}} \textbf{\bibinfo{volume}{437}},
  \bibinfo{pages}{262--272} (\bibinfo{year}{2014}).
\newblock \eprint{1310.1134}.

\bibitem{Harding1999}
\bibinfo{author}{{Harding}, A.~K.}, \bibinfo{author}{{Contopoulos}, I.} \&
  \bibinfo{author}{{Kazanas}, D.}
\newblock \bibinfo{title}{{Magnetar Spin-Down}}.
\newblock \emph{\bibinfo{journal}{\apjl}} \textbf{\bibinfo{volume}{525}},
  \bibinfo{pages}{L125--L128} (\bibinfo{year}{1999}).
\newblock \eprint{astro-ph/9908279}.

\bibitem{Tong2013}
\bibinfo{author}{{Tong}, H.}, \bibinfo{author}{{Xu}, R.~X.},
  \bibinfo{author}{{Song}, L.~M.} \& \bibinfo{author}{{Qiao}, G.~J.}
\newblock \bibinfo{title}{{Wind Braking of Magnetars}}.
\newblock \emph{\bibinfo{journal}{\apj}} \textbf{\bibinfo{volume}{768}},
  \bibinfo{pages}{144} (\bibinfo{year}{2013}).
\newblock \eprint{1205.1626}.

\bibitem{Ramachandran2004}
\bibinfo{author}{{Ramachandran}, R.}, \bibinfo{author}{{Backer}, D.~C.},
  \bibinfo{author}{{Rankin}, J.~M.}, \bibinfo{author}{{Weisberg}, J.~M.} \&
  \bibinfo{author}{{Devine}, K.~E.}
\newblock \bibinfo{title}{{Effect of Quasi-Orthogonal Emission Modes on the
  Rotation Measures of Pulsars}}.
\newblock \emph{\bibinfo{journal}{\apj}} \textbf{\bibinfo{volume}{606}},
  \bibinfo{pages}{1167--1173} (\bibinfo{year}{2004}).
\newblock \eprint{astro-ph/0401534}.

\bibitem{Dai2015}
\bibinfo{author}{{Dai}, S.} \emph{et~al.}
\newblock \bibinfo{title}{{A study of multifrequency polarization pulse
  profiles of millisecond pulsars}}.
\newblock \emph{\bibinfo{journal}{\mnras}} \textbf{\bibinfo{volume}{449}},
  \bibinfo{pages}{3223--3262} (\bibinfo{year}{2015}).
\newblock \eprint{1503.01841}.

\bibitem{Karastergiou2009}
\bibinfo{author}{{Karastergiou}, A.}
\newblock \bibinfo{title}{{The complex polarization angles of radio pulsars:
  orthogonal jumps and interstellar scattering}}.
\newblock \emph{\bibinfo{journal}{\mnras}} \textbf{\bibinfo{volume}{392}},
  \bibinfo{pages}{L60--L64} (\bibinfo{year}{2009}).
\newblock \eprint{0901.1826}.

\bibitem{CHIME2020}
\bibinfo{author}{{CHIME/FRB Collaboration}} \emph{et~al.}
\newblock \bibinfo{title}{{A bright millisecond-duration radio burst from a
  Galactic magnetar}}.
\newblock \emph{\bibinfo{journal}{\nat}} \textbf{\bibinfo{volume}{587}},
  \bibinfo{pages}{54--58} (\bibinfo{year}{2020}).
\newblock \eprint{2005.10324}.

\bibitem{STARE2020}
\bibinfo{author}{{Bochenek}, C.~D.} \emph{et~al.}
\newblock \bibinfo{title}{{A fast radio burst associated with a Galactic
  magnetar}}.
\newblock \emph{\bibinfo{journal}{\nat}} \textbf{\bibinfo{volume}{587}},
  \bibinfo{pages}{59--62} (\bibinfo{year}{2020}).
\newblock \eprint{2005.10828}.

\bibitem{Hilmarsson2021}
\bibinfo{author}{{Hilmarsson}, G.~H.} \emph{et~al.}
\newblock \bibinfo{title}{{Rotation Measure Evolution of the Repeating Fast
  Radio Burst Source FRB 121102}}.
\newblock \emph{\bibinfo{journal}{\apjl}} \textbf{\bibinfo{volume}{908}},
  \bibinfo{pages}{L10} (\bibinfo{year}{2021}).
\newblock \eprint{2009.12135}.

\bibitem{Mckinven2022}
\bibinfo{author}{{Mckinven}, R.} \emph{et~al.}
\newblock \bibinfo{title}{{A Large Scale Magneto-ionic Fluctuation in the Local
  Environment of Periodic Fast Radio Burst Source, FRB 20180916B}}.
\newblock \emph{\bibinfo{journal}{arXiv e-prints}}
  \bibinfo{pages}{arXiv:2205.09221} (\bibinfo{year}{2022}).
\newblock \eprint{2205.09221}.

\bibitem{Desvignes2018}
\bibinfo{author}{{Desvignes}, G.} \emph{et~al.}
\newblock \bibinfo{title}{{Large Magneto-ionic Variations toward the Galactic
  Center Magnetar, PSR J1745-2900}}.
\newblock \emph{\bibinfo{journal}{\apjl}} \textbf{\bibinfo{volume}{852}},
  \bibinfo{pages}{L12} (\bibinfo{year}{2018}).
\newblock \eprint{1711.10323}.

\bibitem{Xu2022}
\bibinfo{author}{{Xu}, H.} \emph{et~al.}
\newblock \bibinfo{title}{{A fast radio burst source at a complex magnetized
  site in a barred galaxy}}.
\newblock \emph{\bibinfo{journal}{\nat}} \textbf{\bibinfo{volume}{609}},
  \bibinfo{pages}{685--688} (\bibinfo{year}{2022}).
\newblock \eprint{2111.11764}.

\bibitem{Feng2022b}
\bibinfo{author}{{Feng}, Y.} \emph{et~al.}
\newblock \bibinfo{title}{{Circular polarization in two active repeating fast
  radio bursts}}.
\newblock \emph{\bibinfo{journal}{Science Bulletin}}
  \textbf{\bibinfo{volume}{67}}, \bibinfo{pages}{2398--2401}
  (\bibinfo{year}{2022}).
\newblock \eprint{2212.05873}.

\bibitem{AnnaThomas2023}
\bibinfo{author}{{Anna-Thomas}, R.} \emph{et~al.}
\newblock \bibinfo{title}{{Magnetic field reversal in the turbulent environment
  around a repeating fast radio burst}}.
\newblock \emph{\bibinfo{journal}{Science}} \textbf{\bibinfo{volume}{380}},
  \bibinfo{pages}{599--603} (\bibinfo{year}{2023}).
\newblock \eprint{2202.11112}.

\bibitem{Feng2022a}
\bibinfo{author}{{Feng}, Y.} \emph{et~al.}
\newblock \bibinfo{title}{{Frequency-dependent polarization of repeating fast
  radio bursts{\textemdash}implications for their origin}}.
\newblock \emph{\bibinfo{journal}{Science}} \textbf{\bibinfo{volume}{375}},
  \bibinfo{pages}{1266--1270} (\bibinfo{year}{2022}).
\newblock \eprint{2202.09601}.

\bibitem{Kumar2023}
\bibinfo{author}{{Kumar}, P.}, \bibinfo{author}{{Shannon}, R.~M.},
  \bibinfo{author}{{Lower}, M.~E.}, \bibinfo{author}{{Deller}, A.~T.} \&
  \bibinfo{author}{{Prochaska}, J.~X.}
\newblock \bibinfo{title}{{Propagation of a fast radio burst through a
  birefringent relativistic plasma}}.
\newblock \emph{\bibinfo{journal}{\prd}} \textbf{\bibinfo{volume}{108}},
  \bibinfo{pages}{043009} (\bibinfo{year}{2023}).
\newblock \eprint{2204.10816}.

\bibitem{Hobbs2020}
\bibinfo{author}{{Hobbs}, G.} \emph{et~al.}
\newblock \bibinfo{title}{{An ultra-wide bandwidth (704 to 4032 MHz) receiver
  for the Parkes radio telescope}}.
\newblock \emph{\bibinfo{journal}{\pasa}} \textbf{\bibinfo{volume}{37}},
  \bibinfo{pages}{e012} (\bibinfo{year}{2020}).
\newblock \eprint{1911.00656}.

\bibitem{Hotan2004}
\bibinfo{author}{{Hotan}, A.~W.}, \bibinfo{author}{{van Straten}, W.} \&
  \bibinfo{author}{{Manchester}, R.~N.}
\newblock \bibinfo{title}{{PSRCHIVE and PSRFITS: An Open Approach to Radio
  Pulsar Data Storage and Analysis}}.
\newblock \emph{\bibinfo{journal}{\pasa}} \textbf{\bibinfo{volume}{21}},
  \bibinfo{pages}{302--309} (\bibinfo{year}{2004}).
\newblock \eprint{astro-ph/0404549}.

\bibitem{vanStraten2011}
\bibinfo{author}{{van Straten}, W.} \& \bibinfo{author}{{Bailes}, M.}
\newblock \bibinfo{title}{{DSPSR: Digital Signal Processing Software for Pulsar
  Astronomy}}.
\newblock \emph{\bibinfo{journal}{Publications of the Astronomical Society of
  Australia}} \textbf{\bibinfo{volume}{28}}, \bibinfo{pages}{1--14}
  (\bibinfo{year}{2011}).
\newblock \eprint{1008.3973}.

\bibitem{Manchester2005}
\bibinfo{author}{{Manchester}, R.~N.}, \bibinfo{author}{{Hobbs}, G.~B.},
  \bibinfo{author}{{Teoh}, A.} \& \bibinfo{author}{{Hobbs}, M.}
\newblock \bibinfo{title}{{The Australia Telescope National Facility Pulsar
  Catalogue}}.
\newblock \emph{\bibinfo{journal}{\aj}} \textbf{\bibinfo{volume}{129}},
  \bibinfo{pages}{1993--2006} (\bibinfo{year}{2005}).
\newblock \eprint{astro-ph/0412641}.

\bibitem{vanStraten2012}
\bibinfo{author}{{van Straten}, W.}, \bibinfo{author}{{Demorest}, P.} \&
  \bibinfo{author}{{Oslowski}, S.}
\newblock \bibinfo{title}{{Pulsar Data Analysis with PSRCHIVE}}.
\newblock \emph{\bibinfo{journal}{Astronomical Research and Technology}}
  \textbf{\bibinfo{volume}{9}}, \bibinfo{pages}{237--256}
  (\bibinfo{year}{2012}).
\newblock \eprint{1205.6276}.

\bibitem{vanStraten2004}
\bibinfo{author}{{van Straten}, W.}
\newblock \bibinfo{title}{{Radio Astronomical Polarimetry and Point-Source
  Calibration}}.
\newblock \emph{\bibinfo{journal}{\apjs}} \textbf{\bibinfo{volume}{152}},
  \bibinfo{pages}{129--135} (\bibinfo{year}{2004}).
\newblock \eprint{astro-ph/0401536}.

\bibitem{vanStraten2010}
\bibinfo{author}{{van Straten}, W.}, \bibinfo{author}{{Manchester}, R.~N.},
  \bibinfo{author}{{Johnston}, S.} \& \bibinfo{author}{{Reynolds}, J.~E.}
\newblock \bibinfo{title}{{PSRCHIVE and PSRFITS: Definition of the Stokes
  Parameters and Instrumental Basis Conventions}}.
\newblock \emph{\bibinfo{journal}{\pasa}} \textbf{\bibinfo{volume}{27}},
  \bibinfo{pages}{104--119} (\bibinfo{year}{2010}).
\newblock \eprint{0912.1662}.

\bibitem{Lower2022ascl}
\bibinfo{author}{{Lower}, M.~E.}, \bibinfo{author}{{Kumar}, P.} \&
  \bibinfo{author}{{Shannon}, R.~M.}
\newblock \bibinfo{title}{{RMNest: Bayesian approach to measuring Faraday
  rotation and conversion in radio signals}}.
\newblock \bibinfo{howpublished}{Astrophysics Source Code Library, record
  ascl:2204.008} (\bibinfo{year}{2022}).
\newblock \eprint{2204.008}.

\bibitem{Han2018}
\bibinfo{author}{{Han}, J.~L.}, \bibinfo{author}{{Manchester}, R.~N.},
  \bibinfo{author}{{van Straten}, W.} \& \bibinfo{author}{{Demorest}, P.}
\newblock \bibinfo{title}{{Pulsar Rotation Measures and Large-scale Magnetic
  Field Reversals in the Galactic Disk}}.
\newblock \emph{\bibinfo{journal}{\apjs}} \textbf{\bibinfo{volume}{234}},
  \bibinfo{pages}{11} (\bibinfo{year}{2018}).
\newblock \eprint{1712.01997}.

\bibitem{Scholz2017}
\bibinfo{author}{{Scholz}, P.} \emph{et~al.}
\newblock \bibinfo{title}{{Spin-down Evolution and Radio Disappearance of the
  Magnetar PSR J1622-4950}}.
\newblock \emph{\bibinfo{journal}{\apj}} \textbf{\bibinfo{volume}{841}},
  \bibinfo{pages}{126} (\bibinfo{year}{2017}).
\newblock \eprint{1705.04899}.

\bibitem{Dewey1985}
\bibinfo{author}{{Dewey}, R.~J.}, \bibinfo{author}{{Taylor}, J.~H.},
  \bibinfo{author}{{Weisberg}, J.~M.} \& \bibinfo{author}{{Stokes}, G.~H.}
\newblock \bibinfo{title}{{A search for low-luminosity pulsars.}}
\newblock \emph{\bibinfo{journal}{\apjl}} \textbf{\bibinfo{volume}{294}},
  \bibinfo{pages}{L25--L29} (\bibinfo{year}{1985}).

\bibitem{Caleb2022}
\bibinfo{author}{{Caleb}, M.} \emph{et~al.}
\newblock \bibinfo{title}{{Radio and X-ray observations of giant pulses from
  XTE J1810 - 197}}.
\newblock \emph{\bibinfo{journal}{\mnras}} \textbf{\bibinfo{volume}{510}},
  \bibinfo{pages}{1996--2010} (\bibinfo{year}{2022}).
\newblock \eprint{2111.01641}.

\bibitem{Johnston2017}
\bibinfo{author}{{Johnston}, S.} \& \bibinfo{author}{{Karastergiou}, A.}
\newblock \bibinfo{title}{{Pulsar braking and the $P$-$\dot{P}$ diagram}}.
\newblock \emph{\bibinfo{journal}{\mnras}} \textbf{\bibinfo{volume}{467}},
  \bibinfo{pages}{3493--3499} (\bibinfo{year}{2017}).
\newblock \eprint{1702.03616}.

\bibitem{bilby}
\bibinfo{author}{{Ashton}, G.} \emph{et~al.}
\newblock \bibinfo{title}{{BILBY: A User-friendly Bayesian Inference Library
  for Gravitational-wave Astronomy}}.
\newblock \emph{\bibinfo{journal}{\apjs}} \textbf{\bibinfo{volume}{241}},
  \bibinfo{pages}{27} (\bibinfo{year}{2019}).
\newblock \eprint{1811.02042}.

\bibitem{dynesty}
\bibinfo{author}{{Speagle}, J.~S.}
\newblock \bibinfo{title}{{DYNESTY: a dynamic nested sampling package for
  estimating Bayesian posteriors and evidences}}.
\newblock \emph{\bibinfo{journal}{\mnras}} \textbf{\bibinfo{volume}{493}},
  \bibinfo{pages}{3132--3158} (\bibinfo{year}{2020}).
\newblock \eprint{1904.02180}.

\bibitem{Melrose2010}
\bibinfo{author}{{Melrose}, D.~B.}
\newblock \bibinfo{title}{{Faraday Rotation: Effect of Magnetic Field
  Reversals}}.
\newblock \emph{\bibinfo{journal}{\apj}} \textbf{\bibinfo{volume}{725}},
  \bibinfo{pages}{1600--1606} (\bibinfo{year}{2010}).
\newblock \eprint{1010.3442}.

\end{thebibliography}

\textbf{Correspondence and requests for materials} should be addressed to Marcus Lower, \href{mailto:marcus.lower@csiro.au}{marcus.lower@csiro.au}.

\textbf{Acknowledgements}
We thank the ATNF staff for their efforts in commissioning the Parkes Ultra-Wideband Low receiver and Medusa signal processor, along with Andrew Jameson and Lawrence Toomey for their guidance regarding software tools and data access.
We also thank Andrew Lyne and Mitch Mickaliger for help with obtaining and preparing the Jodrell Bank data.
M.E.L. thanks Lucy Oswald and Stefan Os{\l}owski for discussions on the impact polarization mode addition has on the properties of pulsars.
The Parkes radio telescope (\textit{Murriyang}) is part of the Australia Telescope National Facility (\url{https://ror.org/05qajvd42}) which is funded by the Australian Government for operation as a National Facility managed by CSIRO.
We acknowledge the Wiradjuri people as the traditional owners of the Observatory site.
Pulsar research at the Jodrell Bank Centre for Astrophysics and Jodrell Bank Observatory is supported by a consolidated grant from the UK Science and Technology Facilities Council (STFC).
This paper includes archived data obtained through the CSIRO Data Access Portal (\url{http://data.csiro.au}).
Data reduction, analysis and computations were performed on the OzSTAR national HPC facility, which is funded by Swinburne University of Technology and the National Collaborative Research Infrastructure Strategy (NCRIS).
This research was funded partially by the Australian Government through the Australian Research Council (ARC) grant CE170100004 (OzGrav).
R.M.S. acknowledges support through ARC Discovery Project DP220102305 and ARC Future Fellowship FT190100155.
M.C. acknowledges support of an ARC Discovery Early Career Research Award DE220100819 funded by the Australian Government and the ARC Centre of Excellence for All Sky Astrophysics in 3 Dimensions (ASTRO 3D), through project number CE170100013.
S.D. is the recipient of an ARC Discovery Early Career Award (DE210101738) funded by the Australian Government.
D.L. acknowledges the National Natural Science Foundation of China \#11988101.
K.M.R. acknowledges support from the Vici research programme ``ARGO'' with project number 639.043.815, financed by the Dutch Research Council (NWO).

\textbf{Contributions to the paper}
M.E.L. drafted the manuscript with suggestions from co-authors.
M.E.L. reduced and analyzed the Parkes radio data, and was responsible for development of the Faraday conversion model with input from S.J. and R.M.S. in equal measure.
Both M.L. and D.B.M. provided insights into theoretical interpretation of the results.
B.S. co-organized the JBO observing campaign and data acquisition.
K.M.W., M.C. and P.W. reduced and calibrated the JBO data.
P.W. contributed the associated text regarding JBO data collection and processing.
F.C. is the PI of the Parkes P885 project, contributed to proposal writing and comments on the manuscript.
J.E.R., J.M.S and M.E.L. conducted observations through the P885 project.
S.D. conducted observations under the Parkes P970 project.
A.D.C, D.L. and G.H. conducted observations through the Parkes PX500/501 projects that contributed to the project

\textbf{Competing Interests}
The authors declare no competing interests.


\clearpage
\setcounter{figure}{0}

\begin{figure*}
    \centering
    \includegraphics[width=\linewidth]{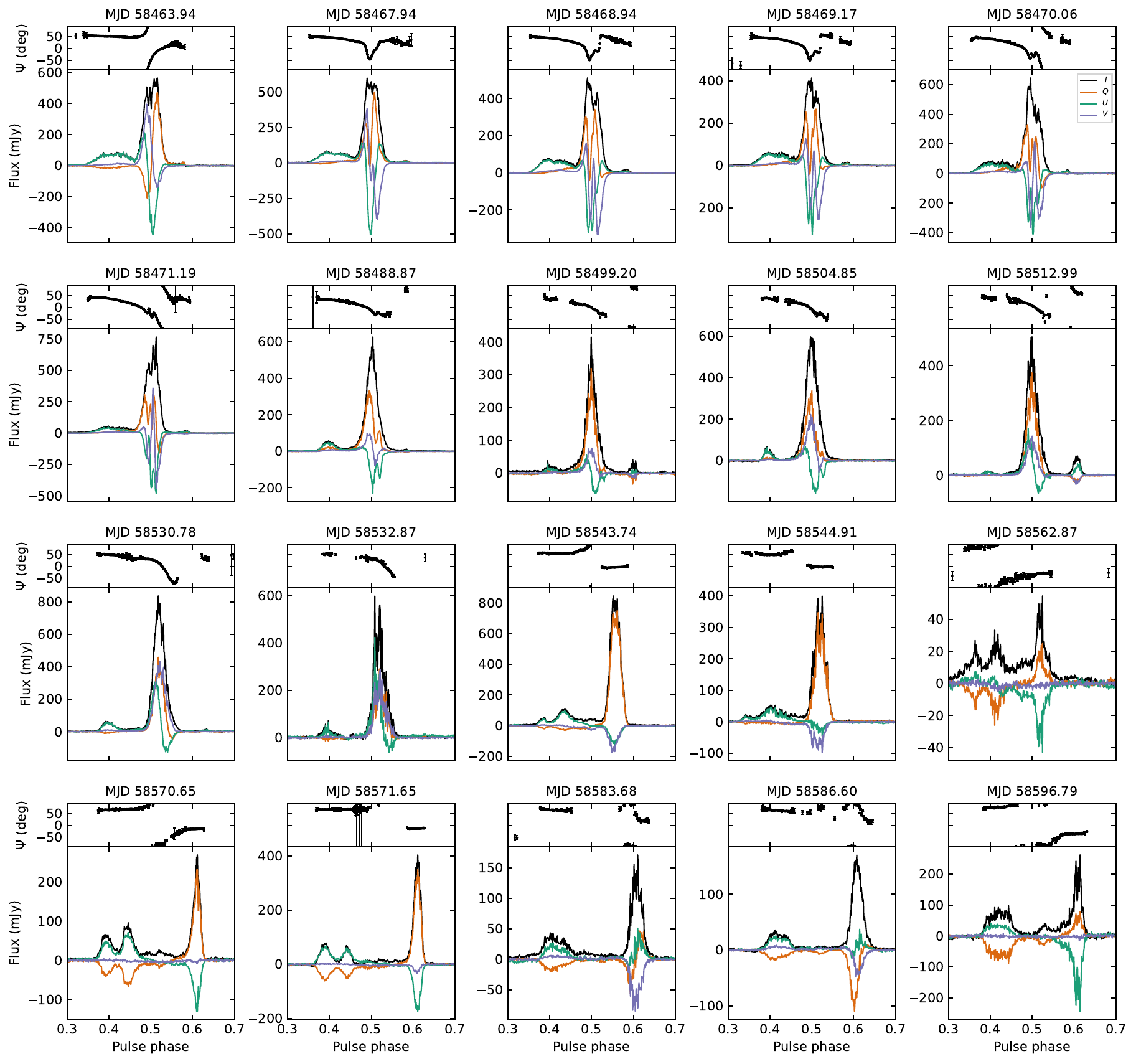}
    \caption{{\bf Evolving polarization profile of XTE~J1810$-$197.} Time- and frequency-averaged polarization profile and linear polarization position angle of XTE J1810$-$197 observed by Parkes from MJD 58463 to 58596. Black, orange, green and purple lines correspond to Stokes $I$, $Q$, $U$ and $V$ respectively.}
    \label{fig:post_stamps}
\end{figure*}

\clearpage

\begin{figure*}
    \centering
    \includegraphics[width=\linewidth]{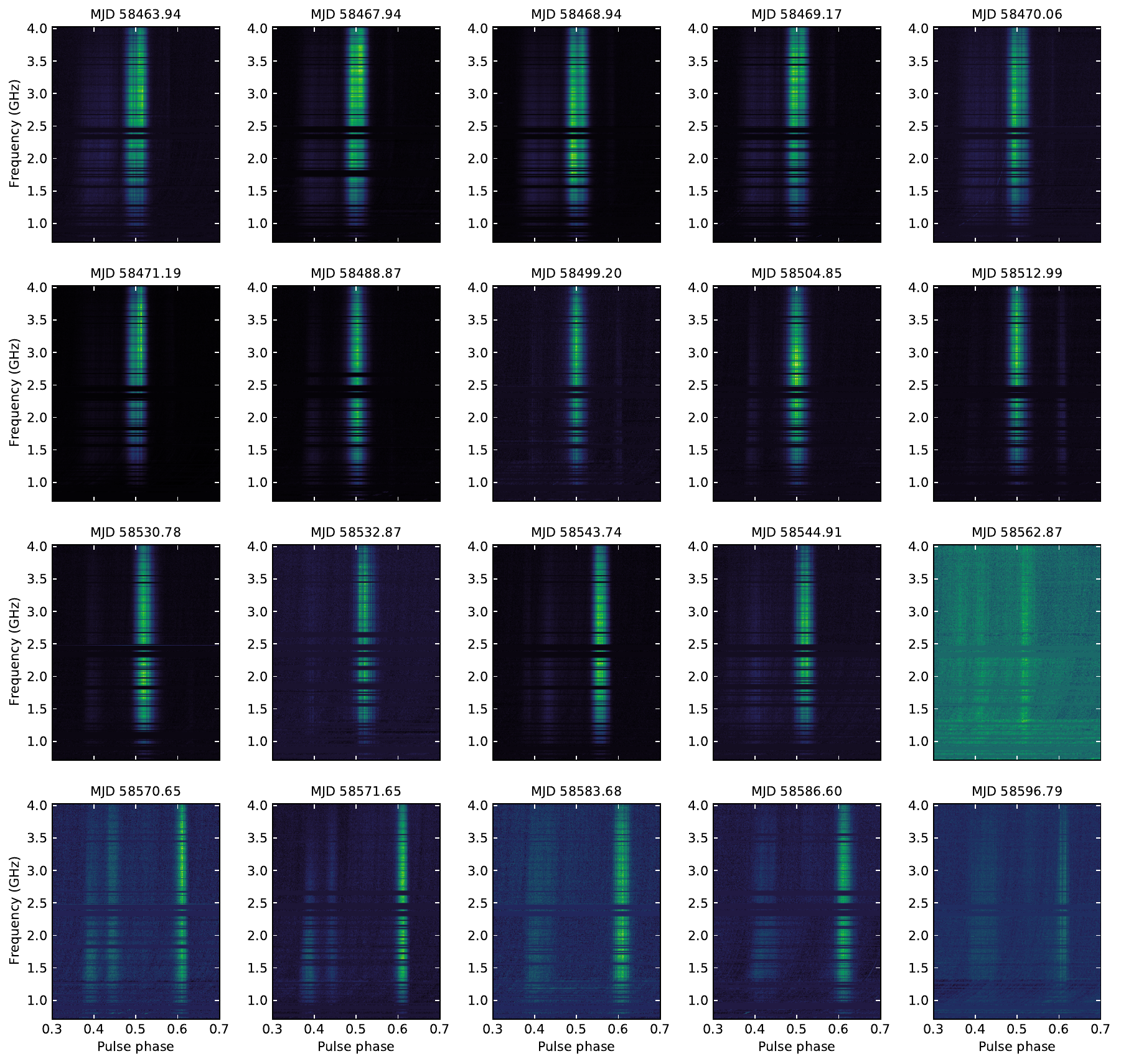}
    \caption{{\bf Total intensity spectrum of XTE~J1810$-$197 over time.} Normalized total intensity of XTE~J1810$-$197 as a function of pulse phase and observing frequency. Horizontal bands in each panel correspond to channels that were excised due to contamination by radio frequency interference.}
    \label{fig:water_stamps}
\end{figure*}

\clearpage

\begin{figure*}[t!]
    \centering
     \begin{subfigure}[b]{0.6\textwidth}
         \centering
         \includegraphics[width=\textwidth]{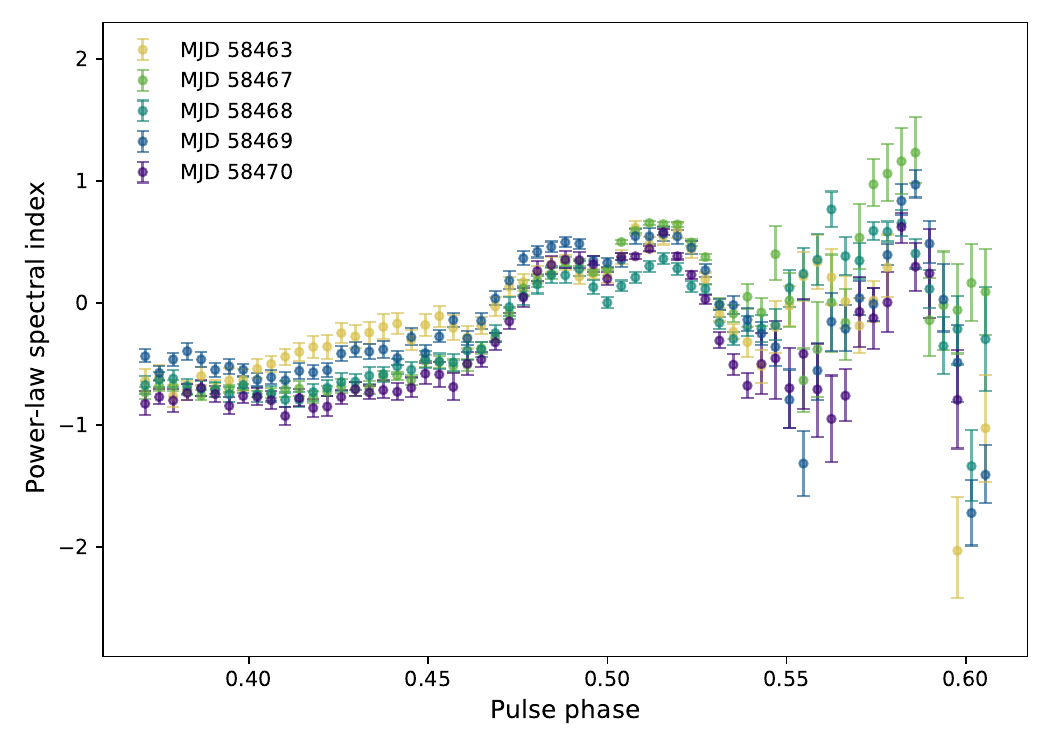}
     \end{subfigure}
     \begin{subfigure}[b]{0.6\textwidth}
         \centering
         \includegraphics[width=\textwidth]{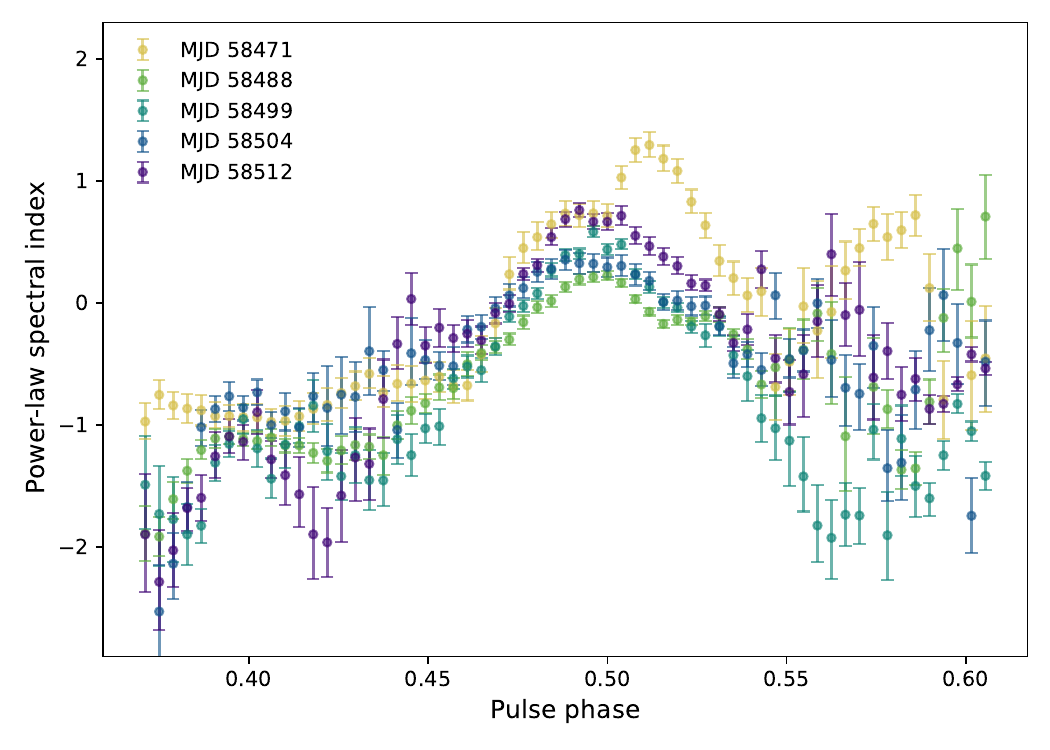}
     \end{subfigure}    
     \caption{{\bf Phase-resolved spectral index evolution of XTE~J1810$-$197.} Median a-posteriori values (points) 68\% confidence intervals (error bars) for the spectral indices measured from power-law fits to the total intensity profile. Results from subsequent epochs are represented by the different colors.}
    \label{fig:spec_index}
\end{figure*}

\clearpage

\begin{figure*}[h!]
    \centering
    \includegraphics[width=0.8\linewidth]{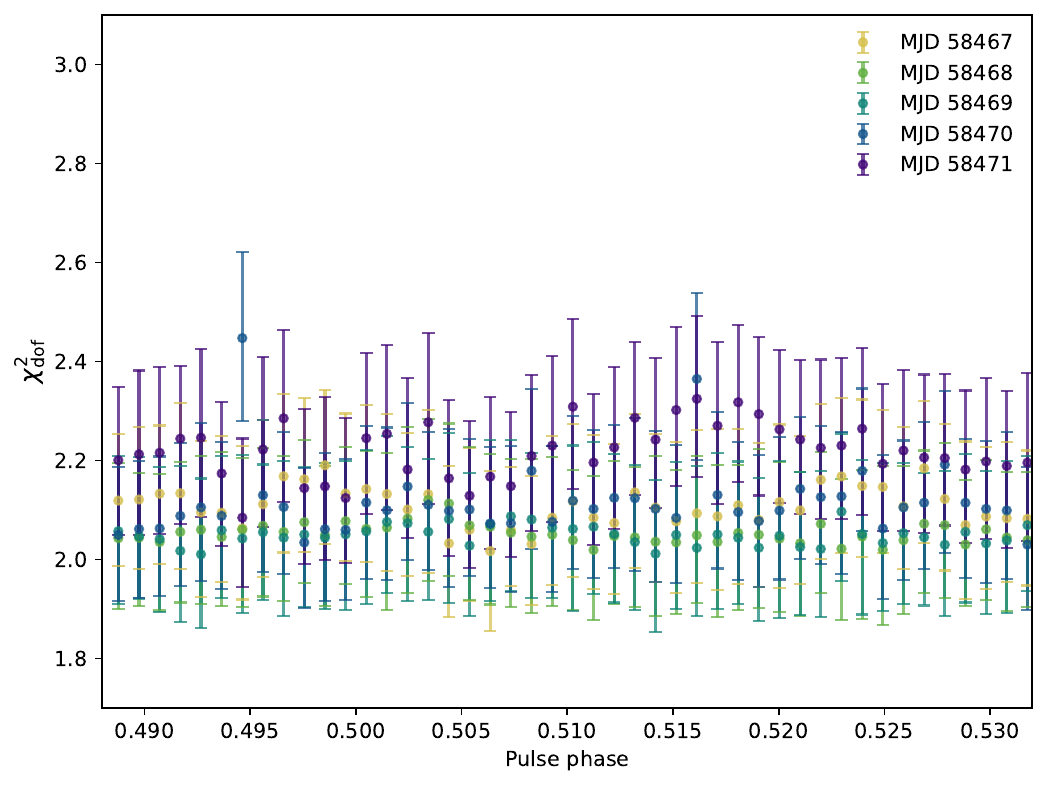}
    \caption{{\bf Reduced chi-square distributions from Faraday conversion fits to XTE~J1810$-$197} Recovered median (points) and 68\% intervals (error bars) of the reduced chi-square statistic ($\chi_{\rm dof}^{2}$) after subtracting 1000 random draws from our phenomenological model fits to the phase-resolved polarization spectra of XTE~J1810$-$197. Each color corresponds to a different epoch.} 
    \label{fig:chi_sq}
\end{figure*}

\clearpage

\begin{figure*}[t!]
    \centering
    \includegraphics[width=0.95\linewidth]{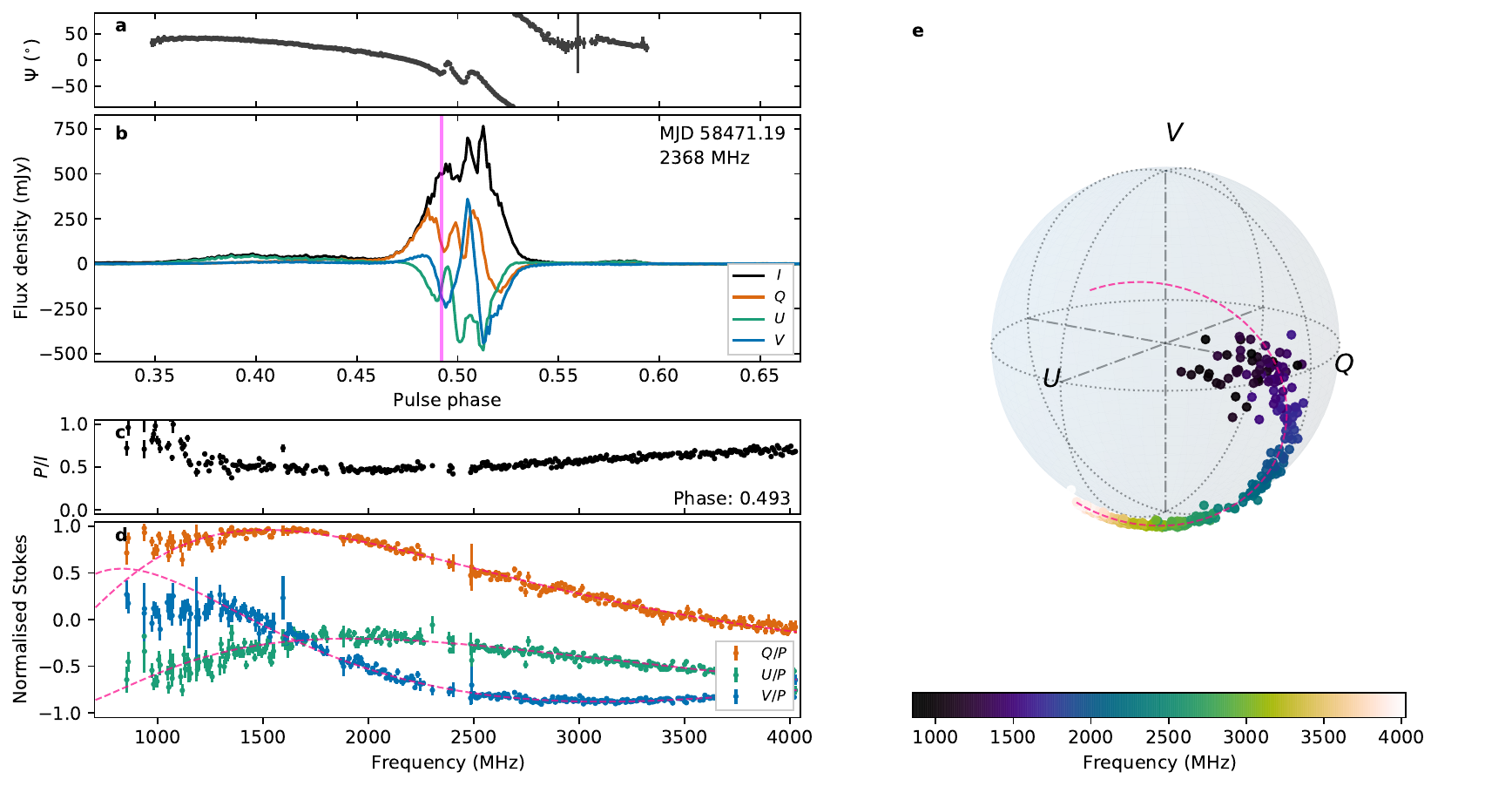}
    \caption{{\bf Example where our Faraday conversion model failed to fit low-frequency variations in XTE~J1810$-$197.} Linear polarization PA and 1-$\sigma$ uncertainties (a) along with the time- and frequency-averaged polarization profile (b) of XTE~J1810$-$197 detected on 19 December 2018, where the black, orange, green and blue lines respectively correspond to the total intensity ($I$), two linear polarizations ($Q$ and $U$) and the circular polarization ($V$). The vertical magenta line indicates the phase bin for which we plot the corresponding time-averaged total polarization fraction (c) and polarization spectra (d) where the error bars correspond to the off-pulse root-mean-square flux. In the latter, Stokes $Q$, $U$ and $V$ were normalized by total polarization ($P$), which traces out an clear frequency-dependent circle on the Poincar\'{e} sphere (e). The dashed magenta lines in panels c and d correspond to the median a-posteriori fit to the data.}
    \label{fig:181219_bin504}
\end{figure*}

\clearpage

\begin{figure*}
     \centering
     \begin{subfigure}[b]{0.41\textwidth}
         \centering
         \includegraphics[width=\textwidth]{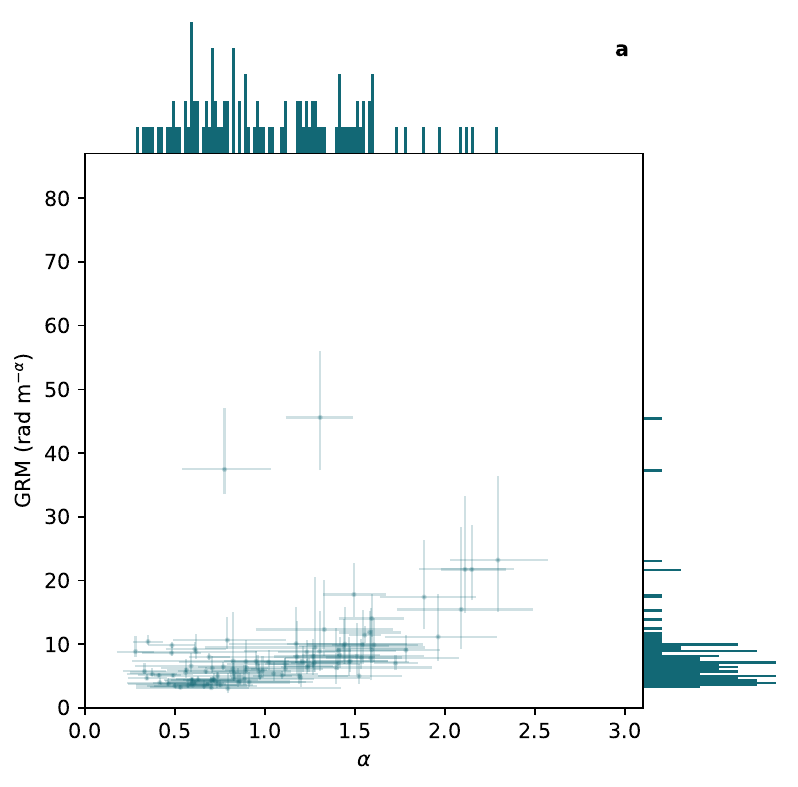}
     \end{subfigure}
     \begin{subfigure}[b]{0.41\textwidth}
         \centering
         \includegraphics[width=\textwidth]{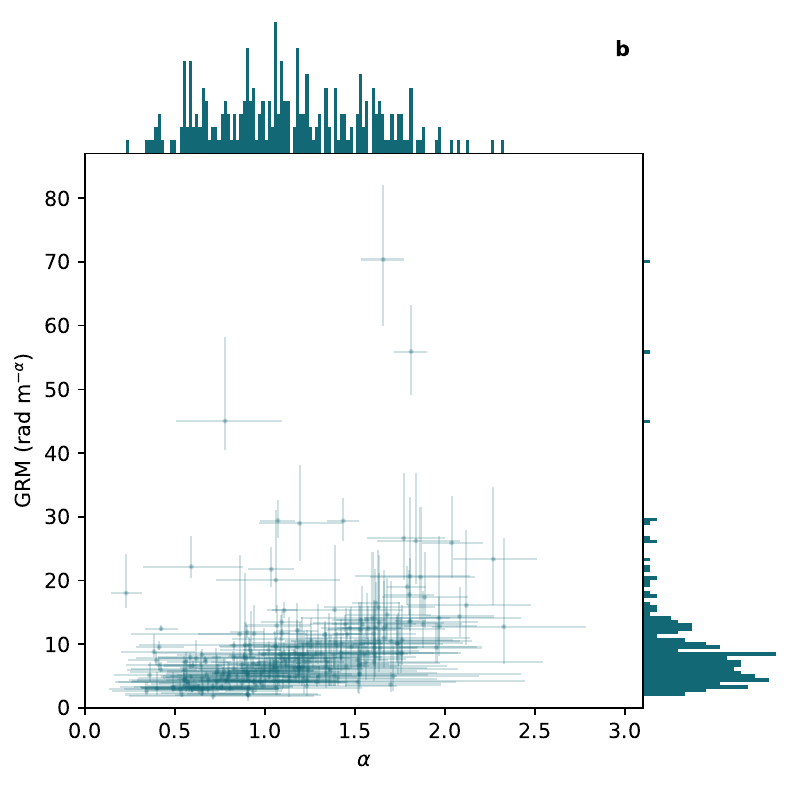}
     \end{subfigure}
     \begin{subfigure}[b]{0.41\textwidth}
         \centering
         \includegraphics[width=\textwidth]{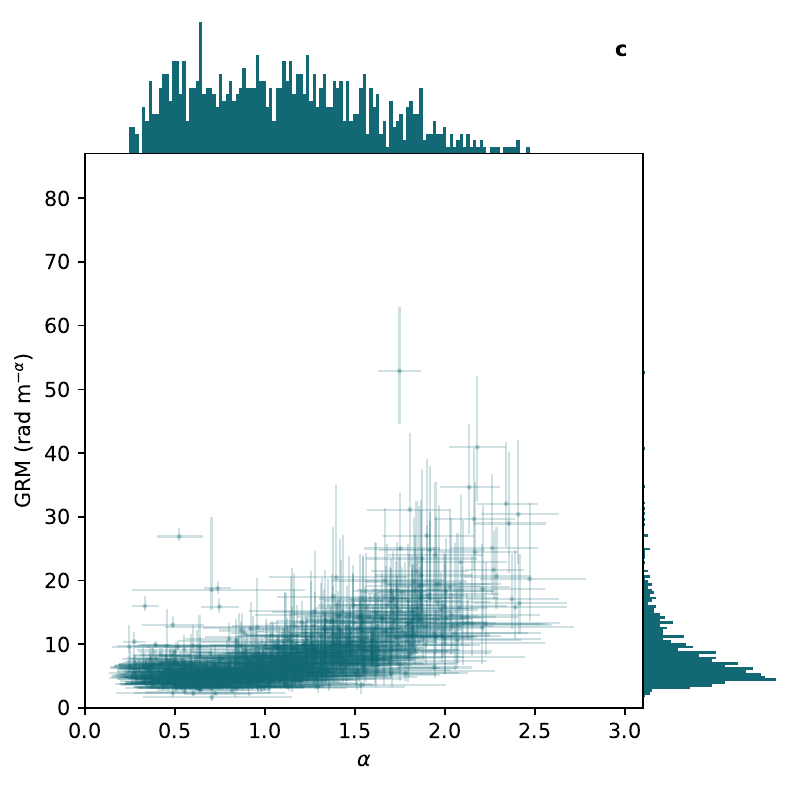}
     \end{subfigure}
     \begin{subfigure}[b]{0.41\textwidth}
         \centering
         \includegraphics[width=\textwidth]{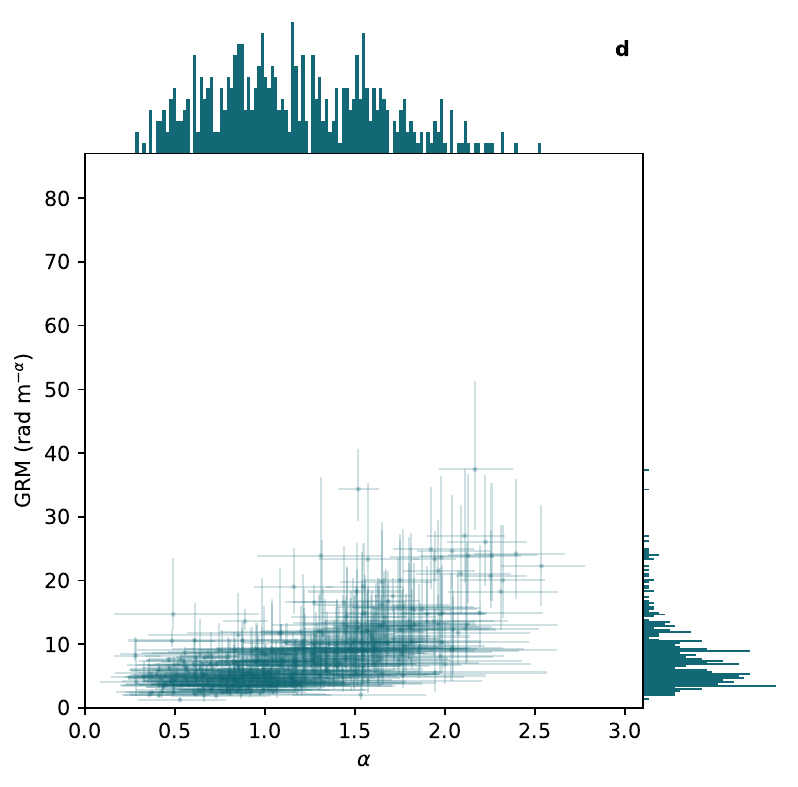}
     \end{subfigure}
     \begin{subfigure}[b]{0.41\textwidth}
         \centering
         \includegraphics[width=\textwidth]{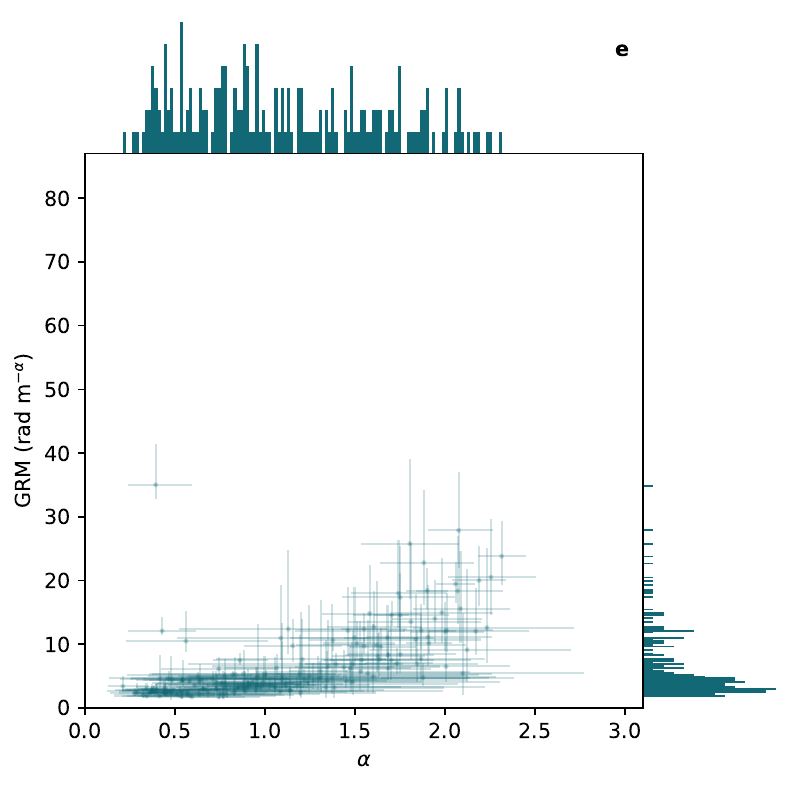}
     \end{subfigure}
    \caption{{\bf Single-pulse generalized rotation measure and wavelength dependencies at different pulse phases.} One- and two-dimensional distributions of the median a-posteriori generalized rotation measure (GRM) and wavelength dependencies ($\alpha)$ from our fits to 2,142 single-pulse polarization spectra detected on MJDs 58468 and 58469. Vertical and horizontal error bars in the two-dimensional distribution correspond to the respective 68\% confidence intervals on our individual measurements. Each subplot contains measurements from pulses phases between 0.483--0.494 (a), 0.494--0.503 (b), 0.503--0.513 (c), 0.513--0.523 (d) and 0.523--0.533 (e).}
    \label{fig:single_dist}
\end{figure*}

\clearpage

\begin{figure*}
     \centering
     \begin{subfigure}[b]{0.45\textwidth}
         \centering
         \includegraphics[width=\textwidth]{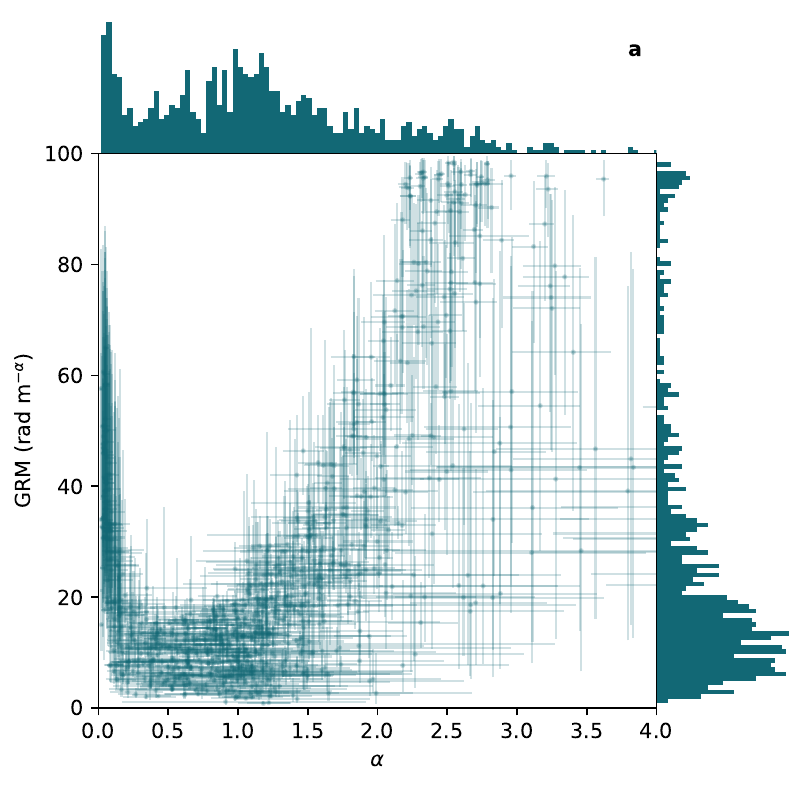}
     \end{subfigure}
     \begin{subfigure}[b]{0.45\textwidth}
         \centering
         \includegraphics[width=\textwidth]{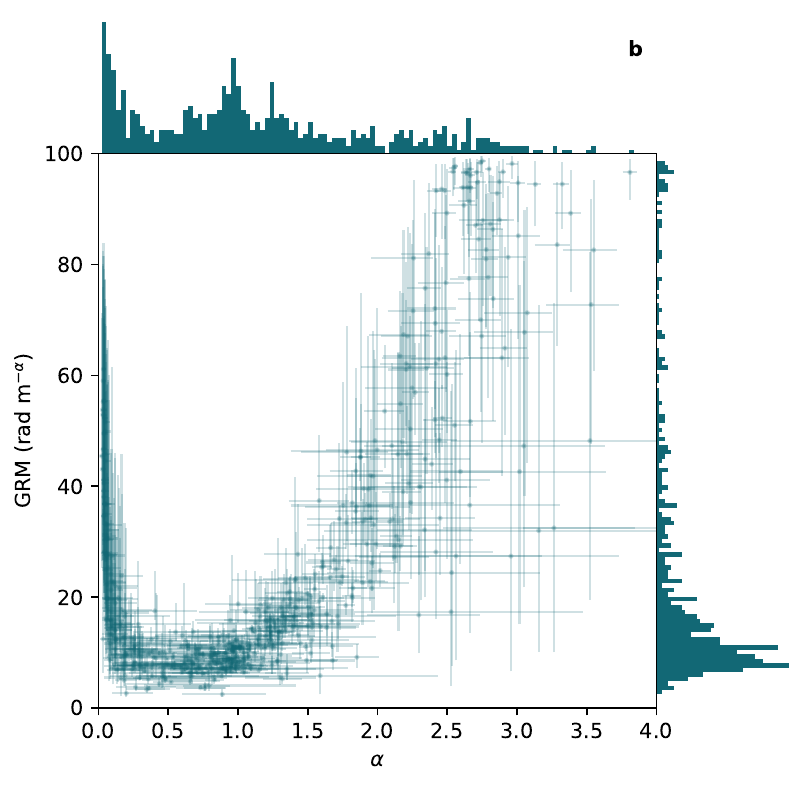}
     \end{subfigure}
     \begin{subfigure}[b]{0.45\textwidth}
         \centering
         \includegraphics[width=\textwidth]{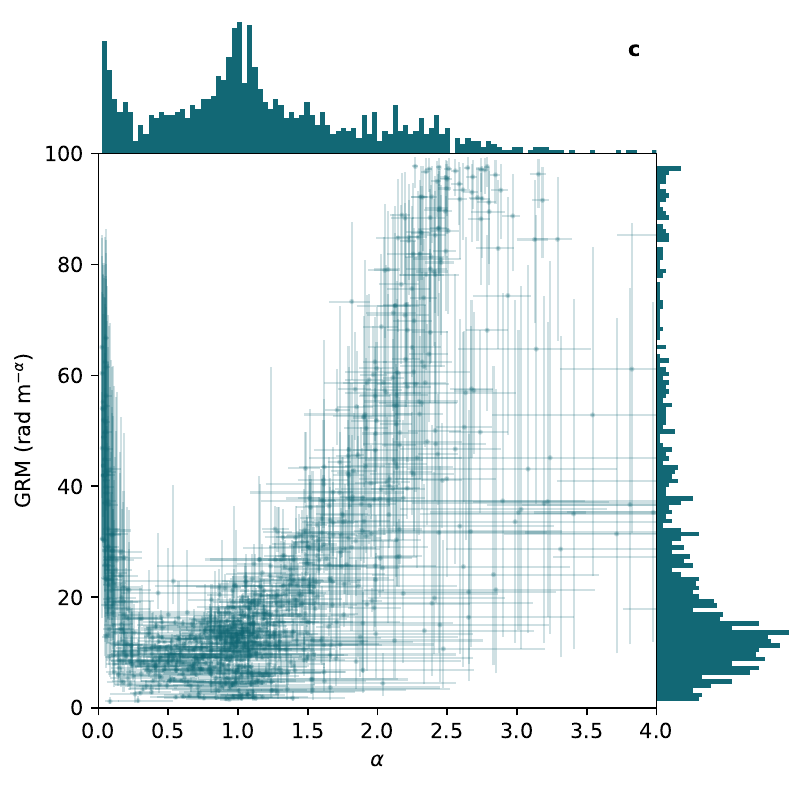}
     \end{subfigure}
     \begin{subfigure}[b]{0.45\textwidth}
         \centering
         \includegraphics[width=\textwidth]{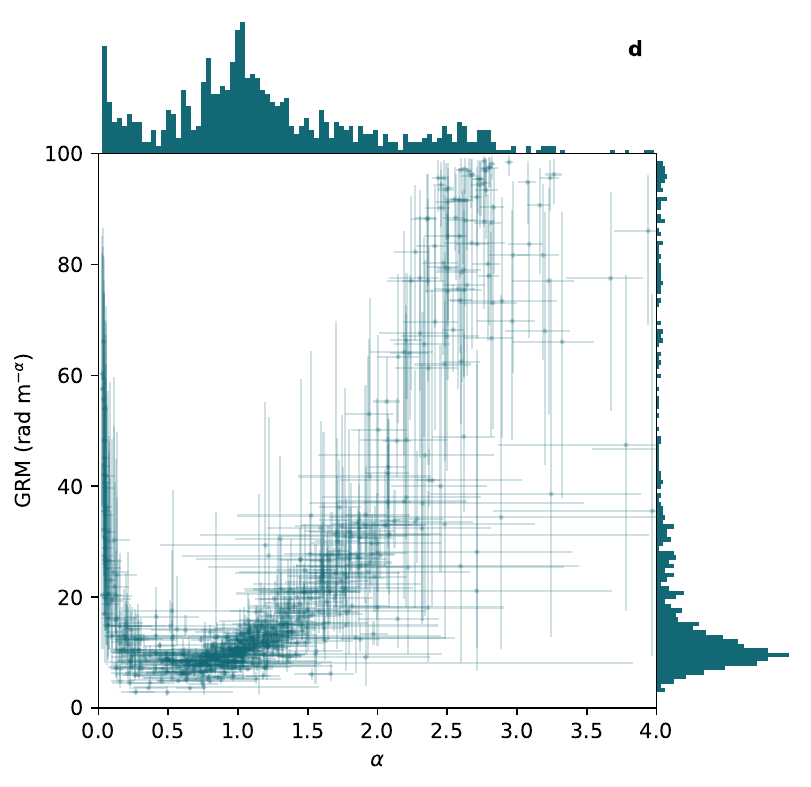}
     \end{subfigure}
        \caption{{\bf Generalized rotation measure and wavelength dependencies inferred from four simulations of propagation through two Faraday screens.} Distributions of the recovered median a-posteriori (with 68\% confidence intervals shown by the errorbars) generalized rotation measure and wavelength dependence inferred from four simulations of incoherent propagation through two Faraday screens. Top row depicts the results from simulations where the rotation about the Stokes $V$ axis was sampled uniformly with the GRM sampled from either a Uniform (a) or Gaussian (b) distribution. Bottom row emulated the effects of two orthogonally polarized modes by sampling the rotation about the $V$-axis from two Gaussian distributions separated by 180-degrees, and with the GRM sampled from either a Uniform (c) or Gaussian (d) distribution.}
        \label{fig:sim_fits}
\end{figure*}

\clearpage

\begin{figure*}
    \centering
    \includegraphics[width=0.9\linewidth]{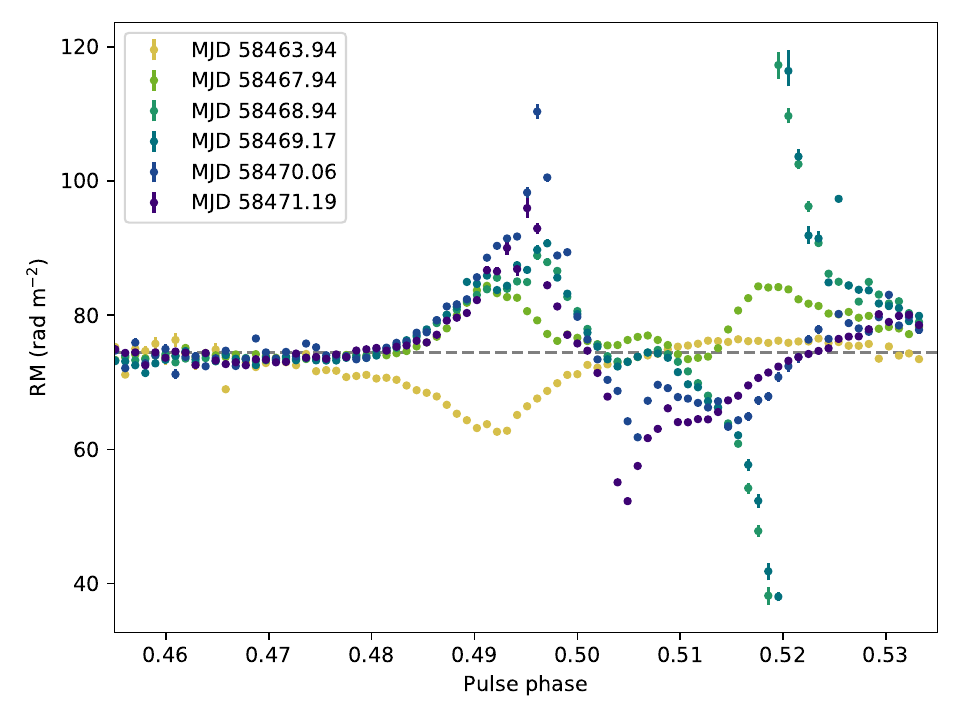}
    \caption{{\bf Phase-resolved rotation measure of XTE~J1810$-$197 in the presence of strong Faraday conversion.} Recovered median a-posteriori and 68\% confidence intervals on the rotation measure of XTE~J1810$-$197 as a function of pulse phase on different epochs. The nominal rotation measure of $74.4$\,rad\,m$^{-2}$ is indicated by the dashed, horizontal line. Strong deviations away from the nominal RM are recovered at phases where intense Faraday conversion is detected. This demonstrates that incorrectly modelled Faraday conversion can potentially account for transient, time-variable rotation measures detected in repeating FRB sources and phase-dependent rotation measures in rotation-powered pulsars.}
    \label{fig:rm_fits}
\end{figure*}

\clearpage
\setcounter{table}{0}

\begin{table*}
\centering
    \caption{Parkes observations of XTE~J1810$-$197.}
    \begin{tabular}{lcccc}
         Date (UTC) & MJD & PID & Length (s) & RM (rad\,m$^{-2}$) \\ 
         \hline
         2018-12-11-22:46:20 & 58463.94 & P970 & 818 & $73.88 \pm 0.07$ \\
         2018-12-15-22:30:08 & 58467.94 & P970 & 1770 & $75.51 \pm 0.05$ \\
         2018-12-16-22:39:51 & 58468.94 & P885 & 3521 & $73.72 \pm 0.03$ \\
         2018-12-17-04:09:33 & 58469.17 & PX500 & 2475 & $72.57 \pm 0.06$ \\
         2018-12-18-01:20:10 & 58470.06 & P970 & 631 & $73.2 \pm 0.1$ \\
         2018-12-19-04:28:52 & 58471.19 & P885 & 1871 & $73.32 \pm 0.07$ \\
         2019-01-05-20:56:23 & 58488.87 & P885 & 1956 & $73.64 \pm 0.08$ \\
         2019-01-16-04:46:20 & 58499.20 & P885 & 882 & $72.8 \pm 0.6$ \\
         2019-01-21-20:17:40 & 58504.85 & P885 & 769 & $75.2 \pm 0.2$ \\
         2019-01-29-23:54:00 & 58512.99 & P885 & 1237 & $76.3 \pm 0.1$ \\
         2019-02-16-18:36:01 & 58530.78 & PX500 & 3629 & $75.0 \pm 0.2$ \\
         2019-02-18-20:57:00 & 58532.87 & P885 & 247 & $78.7 \pm 1.0$ \\
         2019-03-01-17:48:10 & 58543.74 & PX500 & 3040 & $75.00 \pm 0.08$ \\
         2019-03-02-21:50:20 & 58544.91 & P885 & 896 & $75.01 \pm 0.09$ \\
         2019-03-20-20:52:10 & 58562.87 & P885 & 929 & $72.5 \pm 0.5$ \\
         2019-03-28-15:39:17 & 58570.65 & PX500 & 2433 & $76.0 \pm 0.1$ \\
         2019-03-29-15:39:48 & 58571.65 & PX500 & 3042 & $75.21 \pm 0.08$ \\
         2019-04-10-16:20:31 & 58583.68 & P885 & 608 & $73.8 \pm 0.2$ \\
         2019-04-13-14:36:38 & 58586.60 & PX500 & 1420 & $71.8 \pm 0.3$ \\
         2019-04-23-19:06:05 & 58596.79 & P885 & 644 & $75.1 \pm 0.2$ \\
         \hline
    \end{tabular}
    \label{tab:obs}
\end{table*}

\clearpage

\begin{table*}
\centering
    \caption{Parameter distributions used in double Faraday-screen simulations. The injected wavelength dependence was held fixed at $\alpha = 1$ for all simulations. For Uniform distributions, the upper and lower bounds are listed, while the mean and standard deviation are shown for the Normal distributions.}
    \begin{tabular}{lllll}
                         &  Sim. {\bf a}                & Sim. {\bf b}                & Sim. {\bf c}                & Sim. {\bf d} \\ 
         \hline
         GRM             & $\textrm{Uniform}(0, 20)$    & $\textrm{Normal}(10, 2)$    & $\textrm{Uniform}(0, 20)$   & $\textrm{Normal}(10, 2)$    \\  
         $\Psi_{0}$      & $\textrm{Uniform}(-90, 90)$  & $\textrm{Uniform}(-90, 90)$ & $\textrm{Uniform}(-90, 90)$ & $\textrm{Uniform}(-90, 90)$ \\
         $\chi$          & $\textrm{Uniform}(0, 45)$    & $\textrm{Uniform}(0, 45)$   & $\textrm{Uniform}(0, 45)$.  & $\textrm{Uniform}(0, 45)$   \\
         $\varphi_{1}$   & $\textrm{Uniform}(0, 180)$   & $\textrm{Uniform}(0, 180)$  & $\textrm{Normal}(0, 45)$    & $\textrm{Normal}(0, 45)$    \\
         $\varphi_{2}$   & $\textrm{Uniform}(0, 180)$   & $\textrm{Uniform}(0, 180)$  & $\textrm{Normal}(180, 45)$  & $\textrm{Normal}(180, 45)$  \\
         $\vartheta_{1}$ & $\textrm{Uniform}(0, 180)$   & $\textrm{Uniform}(0, 100)$  & $\textrm{Uniform}(0, 90)$   & $\textrm{Uniform}(0, 90)$   \\
         $\vartheta_{2}$ & $\textrm{Uniform}(0, 180)$   & $\textrm{Uniform}(0, 100)$  & $\textrm{Uniform}(0, 90)$  & $\textrm{Uniform}(0, 90)$    \\
         \hline
    \end{tabular}
    \label{tab:sim_priors}
\end{table*}

\end{document}